\documentclass[11pt,draftclsnofoot,onecolumn]{IEEEtran}
\usepackage{graphicx}
\usepackage{epsfig}
\usepackage{amsmath,amssymb,amsfonts,cite}

\def\xv{{\mathbf x}}
\def\yv{{\mathbf y}}

\def\Xv{{\mathbf X}}
\def\Yv{{\mathbf Y}}

\def\Pr{\mathrm{Pr}}

\def\Rc{\mathcal R}
\def\Xc{\mathcal X}
\def\Yc{\mathcal Y}
\def\Wc{\mathcal W}
\def\Cc{\mathcal C}

\def\gf{{\gamma}}

\newtheorem{lemma}{Lemma}
\newtheorem{theorem}{Theorem}
\newtheorem{corollary}{Corollary}
\newtheorem{remark}{Remark}

\IEEEoverridecommandlockouts

\begin{document}

\title{Interference Assisted Secret Communication}

\author{Xiaojun~Tang,~Ruoheng~Liu,~Predrag~Spasojevi\'{c},~and~H.~Vincent~Poor
\thanks{This research was supported by the National Science Foundation
under Grants CNS-06-25637, CCF-07-28208 and CCF-0729142, and in part by the Air Force Office of Scientific Research under Grant  FA9550-08-1-0480. The material in this
paper was presented in part at the IEEE Information Theory Workshop, Porto, Portugal May 5-9 2008, and in part at the IEEE Communication Theory Workshop, Napa, CA, May 10 - 13, 2009.}%
\thanks{X. Tang and P. Spasojevi\'{c} are with Wireless Information Network Laboratory (WINLAB), Department of Electrical and Computer Engineering, Rutgers University,
North Brunswick, NJ 08902, USA (e-mail:
\{xtang,spasojev\}@winlab.rutgers.edu).}%
\thanks{R. Liu and H. V. Poor are with Department of Electrical Engineering, Princeton
University, Princeton, NJ 08544, USA (email:
\{rliu,poor\}@princeton.edu).}}

\maketitle

\begin{abstract}
Wireless communication is susceptible to eavesdropping attacks because of its broadcast nature.
This paper illustrates how interference can be used to counter eavesdropping and assist
secrecy. In particular, a wire-tap channel with a helping interferer (WT-HI) is considered. Here, a transmitter sends
a confidential message to its intended receiver in the presence of a
passive eavesdropper and with the help of an independent interferer.
The interferer, which does not know the confidential message, helps
in ensuring the secrecy of the message by sending an independent
signal. An achievable secrecy rate and several computable outer bounds
on the secrecy capacity of the WT-HI are given for both discrete
memoryless and Gaussian channels.
\end{abstract}


\section{Introduction}\label{sec:intro}

Broadcast and superposition are two fundamental properties of the wireless medium. Due to its broadcast nature, wireless transmissions can be received by multiple receivers with possibly different signal strengths. This property makes wireless communications susceptible to eavesdropping. Due to the superposition property of the wireless medium, a receiver observes a superposition of multiple simultaneous transmissions resulting in interference. This paper illustrates how one can pit the superposition property of the wireless medium against eavesdropping by using interference to assist secrecy.

Our approach follows Wyner's seminal work on the wire-tap channel \cite{Wyner:BSTJ:75}, in which a single source-destination communication is eavesdropped upon via a degraded channel.  Wyner's formulation was generalized by Csisz{\'{a}}r and K{\"{o}}rner who studied general broadcast channels \cite{Csiszar:IT:78}. The Gaussian wire-tap channel was considered in \cite{Leung-Yan-Cheong:IT:78}. In these models, it is desirable to minimize the leakage of information to the eavesdropper. The level of ignorance of the eavesdropper with respect to the confidential messages is measured by the equivocation rate. Perfect secrecy requires that the equivocation rate is asymptotically equal to the message rate, and the maximal achievable rate with perfect secrecy is the secrecy capacity. The central idea of \cite{Wyner:BSTJ:75,Csiszar:IT:78,Leung-Yan-Cheong:IT:78} is that the transmitter can use stochastic encoding to introduce randomness to preserve secrecy. In this paper, we study the problem in which a transmitter sends confidential messages to the intended receiver with the help of an interferer, in the presence of a passive eavesdropper. The difference between this model and Csisz{\'{a}}r and K{\"{o}}rner's model is that there is an additional transmitter, who functions as an interferer without any knowledge of the actual confidential message sent by the primary transmitter. We call this model the wire-tap channel with a helping interferer (WT-HI). The external transmitter provides additional randomization to increase the secrecy level of the primary transmission. We choose the transmission schemes at both the interferer and the legitimate transmitter to enhance the secrecy rate.

To understand the effects of interference in wireless transmissions, the interference channel (IC) has been extensively studied. The capacity region of interference channels remains an open problem, except for some special cases including the strong/very strong interference regimes \cite{Carleial:IT:75} and \cite{Sato:IT:81}. The best achievable rate region so far was proposed by Han and Kobayashi \cite{Han:IT:81}. Several outer bounds for the Gaussian IC with weak interference were proposed in \cite{Sato:IT:78,Costa:IT:85,Kramer:IT:04,Etkin:IT:08}, and more recently a new outer bound was proposed independently in \cite{Motahari:IT:09,Annapureddy:ITA:08,Shang:IT:09} to obtain the sum-capacity in a very weak interference regime.

The secrecy capacity of interference channels remains even more elusive. In \cite{Liu:IT:08}, an achievable secrecy rate region and an outer bound were proposed for the discrete memoryless interference channel with confidential messages (IC-CM). An achievable secrecy rate was also proposed for Gaussian IC-CMs. In \cite{Liang:IT:09}, the secrecy capacity region was found for a special class of cognitive interference channels, in which the cognitive user knows the message sent by the primary user non-causally and the primary user is constrained by using deterministic encoding. Secret communication on interference channels was also studied in \cite{Li:ISIT:08,Yates:ISIT:08,Koyluoglu:IT:08}. In \cite{Li:ISIT:08}, an outer bound on the secrecy capacity region of a class of one-sided interference channels was presented. In \cite{Yates:ISIT:08}, the robust-secrecy capacity was defined and characterized for a special deterministic interference channel. In \cite{Koyluoglu:IT:08}, an interference alignment scheme was proposed for the purpose of preserving secrecy.

The information-theoretic secrecy approach has also been applied to study other various multi-user channel models such as the multiple access channel with confidential messages (MAC-CM) \cite{Liang:IT:06,Liu:ISIT:06}, the multiple access wire-tap channel (MAC-WT) \cite{Tekin:IT:07}, and the relay-eavesdropper channel (REC) \cite{Lai:IT:06,Yusel:CISS:07}. We refer the reader to \cite{Liang:ITS:08} for a recent survey of the research progress in this area.

The main contributions of this paper are summarized as follows: First, for general discrete memoryless WT-HI models, we consider all possible interference patterns and design the corresponding achievable coding scheme at the legitimate transmitter based on the coding rate of the interference codebook. We propose an achievable secrecy rate for this channel by optimizing the coding schemes at both the interferer and the legitimate transmitter. Second, for a Gaussian WT-HI, we provide an achievable secrecy rate based on Gaussian codebooks and describe a power policy to optimize the secrecy rate. Our results show that the interferer can increase the secrecy level, and that a positive secrecy rate can be achieved even when the source-destination channel is worse than the source-eavesdropper channel. An important example for the Gaussian case is that in which the interferer has a better channel to the intended receiver than to the eavesdropper. Here, the interferer can send a (random) codeword at a rate that ensures that it can be decoded and subtracted from the received signal by the intended receiver, but cannot be decoded by the eavesdropper. Hence, only the eavesdropper is interfered with and the secrecy level of the confidential message can be increased. In particular, when the interferer-receiver channel is good enough and the power used at the transmitters is unconstrained, the achieved secrecy rate for the Gaussian WT-HI is equal to the secrecy rate achieved when the message is given to the helper secrectly and the helper resends the message. This is particularly interesting because we do not assume that there is a secret transmitter-interferer channel (which would enable the interferer to relay the transmission). Finally, we provide several computable upper bounds on the secrecy capacity of the Gaussian WT-HI model. Each of them can be a better upper bound than others under certain channel and power conditions. For some special cases, the best upper bound is quite close to the achievable secrecy rate.

The WT-HI model has been studied in part within the context of the REC\cite{Lai:IT:06}, MAC-WT \cite{Tekin:IT:07} and IC-CM \cite{Liu:IT:08} models. Our achievable scheme can be considered to be a generalization of the schemes proposed previously. In the cooperative jamming \cite{Tekin:IT:07} scheme or the artificial noise scheme in \cite{Liu:IT:08} (both proposed for Gaussian channels), the helper generates an independent (Gaussian) noise. This scheme does not employ any structure in the transmitted signal and can be considered as a special case of our scheme when the coding rate of the interference codebook is large (infinity). The noise forwarding scheme in \cite{Lai:IT:06} requires that the interferer's codewords can always be decoded by the intended receiver, which can be considered as a special case of our scheme when the coding rate of the interference codebook is lower than a certain rate such that the intended receiver can decode the interference first. By taking a holistic view, we obtain a number of new insights.

The remainder of the paper is organized as follows. Section~\ref{sec:model} describes the system model for the WT-HI. Section \ref{sec:result} states an achievable secrecy rate and a Sato-type upper bound for general discrete memoryless channels. Section \ref{sec:Gaussian} studies a Gaussian WT-HI model, for which an achievable secrecy rate and a power policy for maximizing the secrecy rate, together with several computable upper bounds on the secrecy capacity are given for the Gaussian WT-HI model. Section \ref{sec:numerical} illustrates the results through some numerical examples. Conclusions are given in Section~\ref{sec:conclusions}.

\section{System Model}\label{sec:model}

\begin{figure}
\centering
  \includegraphics[width=0.8\linewidth]{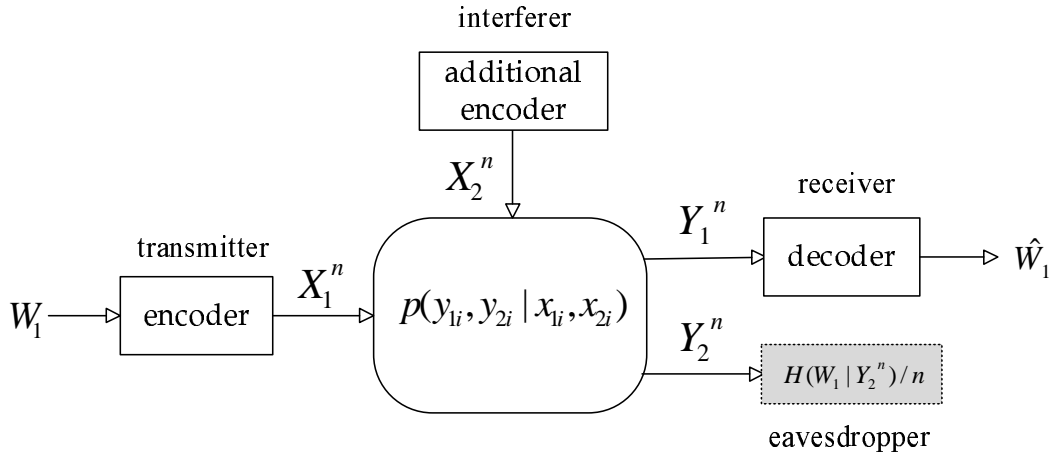}\\
  \caption{A wire-tap channel model with interference: a transmitter wants to send a confidential message $W_1$ to the intended receiver while keeping the message secret with respect to an passive eavesdropper, in the presence of an active interferer.}\label{memorylesschannel}
\end{figure}

As depicted in Fig. \ref{memorylesschannel}, we consider a communication system including a transmitter ($X_1$), an intended receiver ($Y_1$), a helping interferer ($X_2$), and a passive eavesdropper ($Y_2$). The transmitter sends a confidential message $W_1$ to the intended receiver with the help of an \textit{independent} interferer, in the presence of a passive but \textit{intelligent} eavesdropper. We assume that the eavesdropper knows the codebooks of the transmitter and the helper. Furthermore, we assume that the transmitters do not share any common randomness and also that the helper does not know the confidential message $W_1$. As noted above, we refer to this channel as the wire-tap channel with a helping interferer (WT-HI). The channel can be defined by the alphabets $\Xc_1$, $\Xc_2$, $\Yc_1$, $\Yc_2$, and channel transition probability $p(y_1,y_2|x_1,x_2)$ where $x_t\in\Xc_t$ and $y_t\in\Yc_t$, $t=1,2$. The transmitter encodes a confidential message $w_1 \in \Wc_1 = \{1,\dots, M\}$ into $x_1^n$ and sends it to the intended receiver in $n$ channel uses. A stochastic encoder \cite{Csiszar:IT:78} $f_1$ is specified by a matrix of conditional probabilities $f_1(x_{1,k}|w)$, where $x_{1,k} \in \Xc_1$, $w_1 \in \Wc_1$, $\sum_{x_{1,k}}f_1(x_{1,k}|w_1)=1$ for all $k=1,\dots, n$, and $f_1(x_{1,k}|w_1)$ is the probability that the encoder outputs $x_{1,k}$ when message $w_1$ is being sent. The helper generates its output $x_{2,k}$ randomly and can be considered as using another stochastic encoder $f_2$, which is specified by a matrix of probabilities $f_{2}(x_{2,k})$ with $x_{2,k} \in \Xc_{2}$ and $\sum_{x_{2,k}}f_{2}(x_{2,k})=1.$ We assume that $X_1^n$ and $X_2^n$ are independent. Since randomization can increase secrecy, the legitimate transmitter uses stochastic encoding to introduce \textit{randomness}. Additional randomization is provided by the helper and the secrecy can be increased further.

The decoder uses the output sequence $y_1^n$ to compute its estimate $\hat{w}_1$ of $w_1$.
The decoding function is specified by a (deterministic) mapping $g: \Yc_1^n \rightarrow \Wc_1$. The average probability of error is
\begin{equation}\label{pe}
 P_e=\frac{1}{M}\sum_{w=1}^{M}\Pr\left\{g(Y_1^n) \neq w_1 | w_1 ~\mbox{sent}\right\}.
\end{equation}

The secrecy level (level of ignorance of the eavesdropper with respect to the confidential message $w_1$) is measured by the equivocation rate $(1/n)H(W_1|Y_2^n)$.

A secrecy rate $R_s$ is achievable for the WT-HI if, for any $\epsilon>0$, there exists an ($M,n,P_e$) code so that
\begin{equation}\label{ach_def1}
    M \geq 2^{nR_s}, ~ P_e \leq \epsilon
\end{equation}
\begin{equation}\label{ach_def2}
\text{and} \qquad  R_s - \frac{1}{n}H(W_1|Y_2^n) \leq \epsilon \quad
\qquad ~
\end{equation}
for all sufficiently large $n$. The secrecy capacity is the maximum of all achievable secrecy rates.

\section{Discrete Memoryless Channels: Achievable Secrecy Rate and Upper Bound}\label{sec:result}

In this section, we consider the general discrete memoryless WT-HI model. We present an achievable secrecy rate with an outline of its achievable coding scheme. We also present a computable upper bound on the secrecy capacity.

\subsection{Outline of An Achievable Coding Scheme}
\label{outlineachi}

An achievable scheme involves two independent stochastic codebooks. The encoder at the legitimate transmitter uses codebook $\Cc_1(2^{nR_1},2^{nR_{1,s}}, n)$, where $n$ is the codeword length, $2^{nR_1}$ is the size of the codebook, and $2^{nR_{1,s}}$ is the number of confidential messages that $\Cc_1$ can convey ($R_{1,s}\leq R_1$). In addition, the encoder at the interfering helper uses codebook $\Cc_2(2^{nR_2},n)$, where $2^{nR_2}$ is the codebook size.  This codebook can be considered to be the $\Cc_2(2^{nR_2},1, n)$ code where the number of messages that $\Cc_2$ can convey is $1$ (and therefore with zero effective rate).

The random secrecy binning \cite{Wyner:BSTJ:75} technique is applied to $\Cc_1$, so that the $2^{nR_1}$ codewords are randomly grouped into $2^{nR_{1,s}}$ bins each with $2^{n(R_1-R_{1,s})}$ codewords, where each bin represents a message. During the encoding, to send message $w_1 \in W_1$, the encoder selects one codeword uniformly and randomly in the $w$-th bin and sends it to the channel. Meanwhile, the encoder at the interferer randomly selects a codeword in $\Cc_2$ and sends it to the channel.

In the decoding, after receiving $\yv_1$, the intended receiver declares that $\hat{w}_1$ is sent if either of the
following two events occur:
\begin{enumerate}
\item (\emph{separate decoding}): there is only one codeword in $\Cc_1$ that is jointly typical with $\yv_1$ and the bin index of this codeword is $\hat{w}_1$;
\item (\emph{joint decoding}): there is only one pair of codewords in $\Cc_1$ and $\Cc_2$ that are jointly typical with $\yv_1$ and the bin index of the codeword in $\Cc_1$ is $\hat{w}_1$.
\end{enumerate}
The intended receiver declares that a decoding error occurs if neither 1) nor 2) happens. (Please see Appendix \ref{achievability} for the details.)

\subsection{Achievable Rate}

\begin{figure}[htbp]
  \centerline{\hbox{ \hspace{0.01\linewidth}
    \epsfig{file=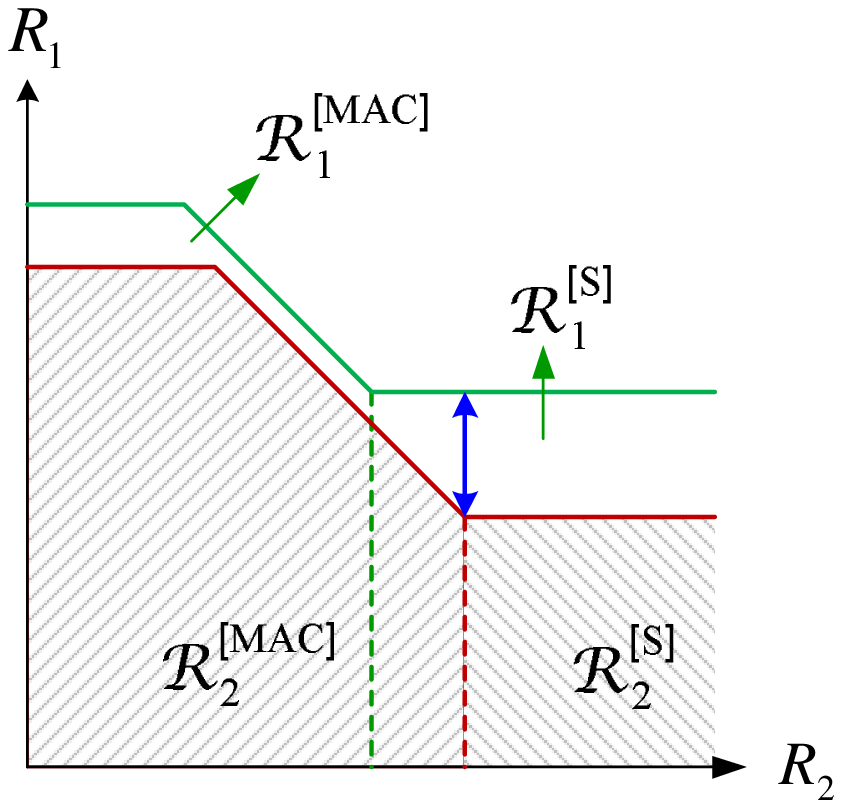, angle=0, width=0.3\linewidth}
    \hspace{0.01\linewidth}
    \epsfig{file=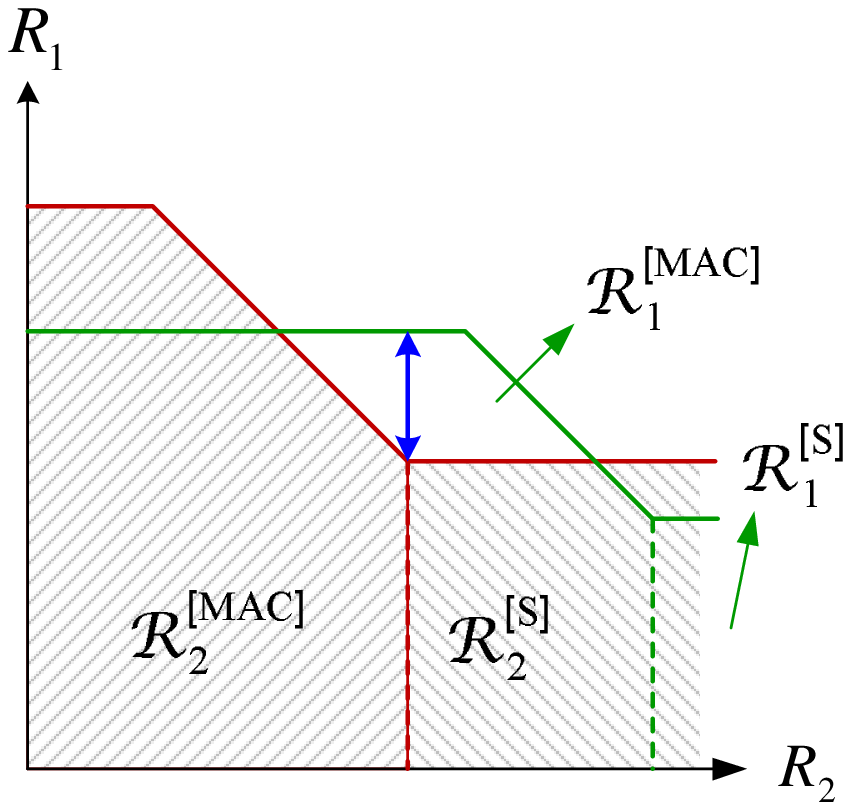, angle=0, width=0.3\linewidth}
    \hspace{0.01\linewidth}
    \epsfig{file=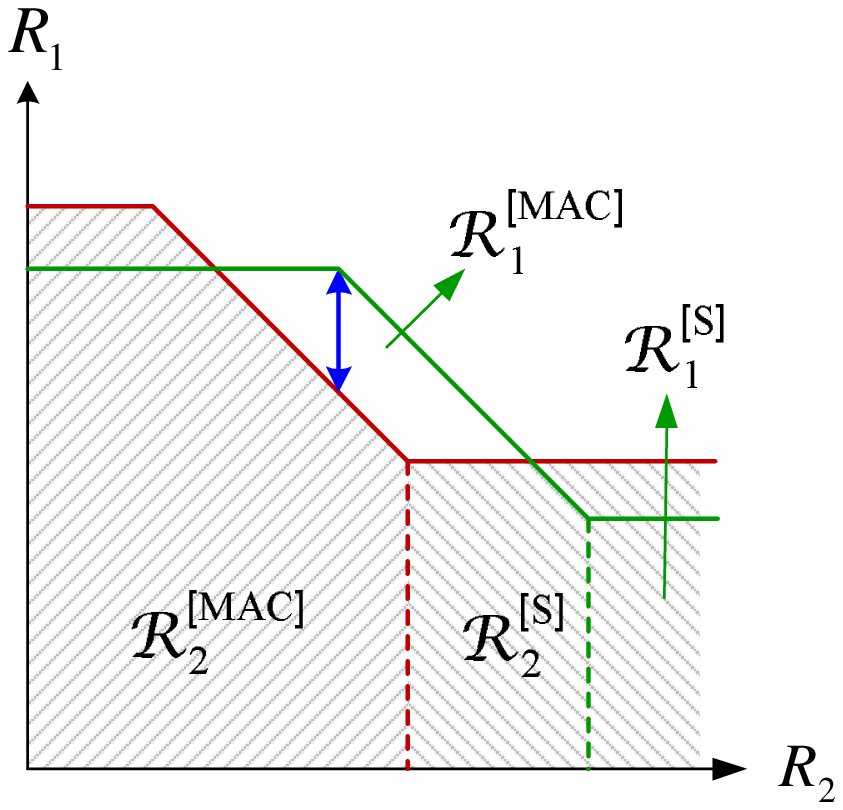, angle=0, width=0.3\linewidth}}
   }
\hbox{\footnotesize \hspace{0.08\linewidth} (a) weak interference
\hspace{0.15\linewidth} (b) strong interference
\hspace{0.15\linewidth} (c) strong interference}
\caption{Code rate $R_1$ versus dummy rate $R_2$ for the intended receiver and eavesdropper.} \label{fig:regions}
\end{figure}

Note that in the achievable scheme, the intended receiver can perform either a joint decoding or a separate decoding. When joint decoding is performed, the intended receiver needs to decode both codewords from $\Cc_1$ and $\Cc_2$. This is essentially a multiple-access channel (MAC) $(\Xc_1,\Xc_2) \rightarrow \Yc_1$.  Hence, We let $\Rc_1^{[\rm MAC]}$ denote the achievable rate region of the MAC $(\Xc_1,\Xc_2) \rightarrow \Yc_1$ defined by
\begin{align}\label{R1_MAC}
\Rc_1^{[\rm MAC]}=\left\{(R_1,R_2)\left|
        \begin{array}{l}
          R_1\ge 0,~R_2\ge 0,\\
          R_1\le I(X_1;Y_1|X_2), \\
          R_2\le I(X_2;Y_1|X_1),\\
          R_1+R_2 \le I(X_1,X_2; Y_1)
        \end{array}
      \right.\right\}.
\end{align}
When separate decoding is performed, the intended receiver does not need to decode $\Cc_2$. Instead, it treats the codewords from $\Cc_2$ as interference. An achievable rate (region) for this separate decoding is given by
\begin{align}\label{R1_S}
\Rc_1^{[\rm S]}=\left\{(R_1,R_2)\left|~
        \begin{array}{l}
          R_1\ge 0,~R_2\ge 0,\\
          R_1\le I(X_1;Y_1),\\
          R_2 >I(X_2;Y_1|X_1)
        \end{array}
      \right.\right\}.
\end{align}
Hence, as shown in Fig.~\ref{fig:regions}, the ``achievable'' rate region in the $R_1$-$R_2$ plane at the receiver is the union of $\Rc^{[\rm MAC]}_1$ and $\Rc^{[\rm S]}_1$.


Similar analysis applies for the eavesdropper as shown in Fig.~\ref{fig:regions}, where $\Rc_2^{[\rm MAC]}$ denotes the region of the MAC $(\Xc_1,\Xc_2) \rightarrow
\Yc_2$:
\begin{align}\label{R2_MAC}
\Rc_2^{[\rm MAC]}=\left\{(R_1,R_2)\left|
        \begin{array}{l}
          R_1\ge 0,~R_2\ge 0,\\
          R_1< I(X_1;Y_2|X_2), \\
          R_2< I(X_2;Y_2|X_1),\\
          R_1+R_2 < I(X_1,X_2; Y_2)
        \end{array}
      \right.\right\},
\end{align}
and $\Rc_2^{[\rm S]}$ is the separate decoding region given by
\begin{align}\label{R2_s}
\Rc_2^{[\rm S]}=\left\{(R_1,R_2)\left|~
        \begin{array}{l}
          R_1\ge 0,~R_2\ge 0,\\
          R_1< I(X_1;Y_2),\\
          R_2 >I(X_2;Y_2|X_1)
        \end{array}
      \right.\right\}.
\end{align}

Our achievable secrecy rate is based on the above definitions of joint and separate decoding regions, and is given in the following theorem.

\begin{theorem} \label{thm:WT-HI}
The following secrecy rate is achievable for the WT-HI:
\begin{align}\label{Rs_WT-HI}
R_s=\max_{\pi, R_1,R_2,R_{1,d}}\left\{R_{1,s}\left|
        \begin{array}{l}
          R_{1,s}+R_{1,d}=R_1,\\
          (R_1,R_2)\in \left\{\Rc_1^{[\rm MAC]} \cup \Rc_1^{[\rm S]}\right\}, \\
          (R_{1,d},R_2) \notin \left\{\Rc_2^{[\rm MAC]} \cup \Rc_2^{[\rm S]}\right\}
        \end{array}
      \right.\right\},
\end{align}
where $\pi$ is the class of distributions that factor as
\begin{equation} \label{eq:dis-IC}
p(x_1)p(x_2)p(y_1,y_2|x_1,x_2).
\end{equation}
\end{theorem}

\begin{remark}
The rate $R_1$ is split as $R_1 = R_{1,s} + R_{1,d}$, where $R_{1,s}$ denotes a secrecy information rate intended by receiver 1 and $R_{1,d}$ represent a redundancy rate sacrificed in order to confuse the eavesdropper. The interferer can help by transmitting dummy information at the rate $R_2$.
\end{remark}

\begin{proof}
The proof consists of error analysis and equivocation computation. It can be found in Appendix \ref{achievability}.

Note that the encoding procedure outlined in Section \ref{outlineachi} involves only one step of binning for $\Cc_1$, but in the proof given in Appendix \ref{achievability}, we assume an additional binning step for $\Cc_1$ (double binning \cite{Liu:IT:08}) and one binning step for $\Cc_2$. However, the additional binning procedure is assumed only for simplifying the proof and is equivalent to the coding procedure described in Section \ref{outlineachi}. More specifically, we do the additional binning for $\Cc_1$ to ensure that some random information can be decoded by the eavesdropper at the rate given by the upper boundary of $\left\{\Rc_2^{[\rm MAC]} \cup \Rc_2^{[\rm S]}\right\}$, if the eavesdropper is interested in decoding the random information when the message $W_1$ is given as side information. This facilitates the technical proof as shown in Appendix \ref{achievability}.

\end{proof}

\subsection{Some Special Cases}

In the following, we consider three typical cases: weak interference/eavesdropping, strong interference/eavesdropping and very strong eavesdropping.

\subsubsection{Weak Interference/Eavesdropping}
This implies that
\begin{align}
&  &         I(X_1;Y_1|X_2)&\ge I(X_1;Y_2|X_2)&\notag\\
&\text{and}& I(X_2;Y_2|X_1)&\ge I(X_2;Y_1|X_1) & \label{eq:IC-W}
\end{align}
for all product distributions on the input $X_1$ and $X_2$. This case is illustrated by Fig. \ref{fig:regions}.(a). Let
\begin{align}
&  &   \Delta_1&=I(X_1;Y_1|X_2)-I(X_1;Y_2|X_2)&\\
&\text{and}& \Delta_2&=I(X_1;Y_1)-I(X_1;Y_2). &
\end{align}
The achievable secrecy rate can be increased by the help from the interferer when $\Delta_1\le\Delta_2$.
The interferer generates an ``artificial noise'' with the dummy rate $R_2>I(X_2;Y_2|X_1)$ so that neither the receiver nor the eavesdropper can decode $\Cc_2$. On the other hand, when $\Delta_1>\Delta_2$, the interferer ``facilitates'' the transmitter by properly choosing the signal $X_2$ to maximize $\Delta_1$. Therefore, in the weak interference case, the intended receiver performs a separate decoding of $\Cc_1$. The achievable secrecy rate can be summarized as
\begin{align*}
R_s = \max_{\pi}\left\{ \max \left(\Delta_1,\Delta_2\right)\right\}.
\end{align*}

\subsubsection{Strong Interference/Eavesdropping}
This implies that
\begin{align}
&  &        I(X_1;Y_1|X_2)&\le I(X_1;Y_2|X_2) &\notag\\
&\text{and}& I(X_2;Y_2|X_1)& \le I(X_2;Y_1|X_1) &\label{eq:IC-S}
\end{align}
for all product distributions on the input $X_1$ and $X_2$. This case is illustrated by Fig. \ref{fig:regions}.(b) and Fig. \ref{fig:regions}.(c). This condition implies that, without the interferer, the channel
$\Xc_1\rightarrow \Yc_2$ is more capable than the channel $\Xc_1\rightarrow \Yc_1$ and, hence, the achievable secrecy rate may
be $0$.


However, we may achieve a positive secrecy rate with the help of the interferer. Here we choose the rate pair $(R_1,R_2)\in \Rc_1^{[\rm MAC]}$ so that the intended
receiver can first decode $\Cc_2$ and then $\Cc_1$. Therefore, in this case, the intended receiver performs joint decoding. Moreover, the dummy rate pair satisfies
$$(R_{1,d},R_2)\notin \left\{\Rc_2^{[\rm MAC]}\cup \Rc_2^{[\rm S]}\right\};$$
i.e., we provide enough randomness to confuse the eavesdropper. Hence, for strong interference, the achievable secrecy rate can be simplified as
\begin{align*}
R_s = \max_{\pi}\left\{ \min \left[
               \begin{array}{l}
                I(X_1,X_2;Y_1)-I(X_1,X_2;Y_2),\\
                I(X_1;Y_1|X_2)-I(X_1;Y_2)
               \end{array}
             \right]\right\}^{+}.
\end{align*}

\subsubsection{Very Strong Eavesdropping}
In this case,
\begin{align}
I(X_1;Y_2)\ge I(X_1;Y_1|X_2)
\end{align}
for all product distributions on the input $X_1$ and $X_2$. We cannot obtain any positive secrecy rate by using the proposed scheme.

The secrecy rate may be increased by using the channel prefixing technique in \cite[Lemma~4]{Csiszar:IT:78}, as shown in the following corollary.
\begin{corollary} \label{thm:WT-HI}
If $X_1$ and $X_2$ in $\Rc_t^{[\rm MAC]}$ and $\Rc_t^{[\rm S]}, t=1,2$, defined by (\ref{R1_MAC})-(\ref{R2_s}), are replaced with random variables $V_1$ and $V_2$, respectively, and the input distribution $\pi$ in (\ref{Rs_WT-HI}) is replaced with $\pi'$, where $\pi'$ is the class of distributions that factor as
\begin{equation}
p(v_1,v_2,x_1,x_2,y_1,y_2)=p(v_1)p(v_2)p(x_1|v_1)p(x_2|v_2)p(y_1,y_2|x_1,x_2),
\end{equation}
then the secrecy rate given by (\ref{Rs_WT-HI}) is achievable.
\end{corollary}

However, we do not follow the prefixing approach in this paper to avoid the intractability of its evaluation.

\subsection{A Sato-type Upper Bound}\label{sec:results:satoupper}

A trivial upper bound on the secrecy capacity is the (main channel) capacity without secrecy constraint. That is
\begin{equation}\label{eq:mainup}
    R_s \leq \max_{P_{X_1},P_{X_2}}I(X_1;Y_1|X_2).
\end{equation}

Here, another computable upper bound for a general WT-HI is a Sato-type upper bound.

\begin{theorem}\label{Souterbound}
The secrecy capacity of the WT-HI satisfies
\begin{equation}\label{sato}
    R_s \leq \min_{P_{\tilde{Y}_1, \tilde{Y}_2|
    X_1,X_2}}\max_{P_{X_1},P_{X_2}}I(X_1,X_2;\tilde{Y}_1|\tilde{Y}_2),
\end{equation}
where $\tilde{Y}_1$ and $\tilde{Y}_2$ are outputs of a discrete memoryless channel characterized by
$P_{\tilde{Y}_1,\tilde{Y}_2|X_1,X_2}$  whose marginal distributions satisfy
\begin{equation}\label{eqalmar}
P_{\tilde{Y}_j|X_1,X_2}(y_j|x_1,x_2) = P_{Y_j|X_1,X_2}(y_j|x_1,x_2),
\end{equation}
for $j=1,2$ and all $y_1$, $y_2$, $x_1$, and $x_2$.
\end{theorem}

\begin{proof}
The proof can be found in Appendix \ref{App-Souter}.
\end{proof}

\begin{remark}
The upper bound assumes that a genie gives the eavesdropper's signal $\tilde{Y}_2$ to the intended receiver as side information for decoding message $W$. Since the eavesdropper's signal $\tilde{Y}_2$ is always a degraded version of the combined signal $(\tilde{Y}_1,\tilde{Y}_2)$, the wire-tap channel result \cite{Wyner:BSTJ:75} can therefore be used.
\end{remark}

\begin{remark}
The upper bound is tight for the degraded WT-HI which satisfies
\begin{equation}\label{degraded}
    P_{Y_2|X_1,X_2}(y_2|x_1,x_2)=\sum_{y_1}P_{Y_1|X_1,X_2}(y_1|x_1,x_2)P_{Y_2|Y_1}(y_2|y_1).
\end{equation}
In the degraded case, the side information $\tilde{Y}_2$ does not benefit the decoding at the intended receiver.
\end{remark}

\section{Gaussian Channels}\label{sec:Gaussian}

In this section, we consider a discrete memoryless Gaussian channel,
for which the channel outputs at the intended receiver and the eavesdropper can be written
as
\begin{align}\label{signalYZ}
  &{~~~~~~} Y_{1,k} = X_{1,k} +\sqrt{b}X_{2,k} + Z_{1,k}, \nonumber \\
  &\mbox{and~~} Y_{2,k} = \sqrt{a}X_{1,k} + X_{2,k} + Z_{2,k},
\end{align}
for $k=1, \dots, n$, where $\{Z_{1,k}\}$ and $\{Z_{2,k}\}$ are sequences of independent and identically distributed (i.i.d.) zero-mean Gaussian noise (real) variables with
unit variances. The channel inputs $X_{1,k}$ and $X_{2,k}$ satisfy average block power constraints of the form
\begin{equation}\label{power}
  \frac{1}{n}\sum_{k=1}^{n}E[X_{1,k}^2] \leq \bar{P}_1 ~~\mbox{and}~~ \frac{1}{n}\sum_{k=1}^{n}E[X_{2,k}^2] \leq
  \bar{P}_2.
\end{equation}
We note that the channel described by (\ref{signalYZ}) satisfies the degradedness condition as defined by (\ref{degraded}) if $ab=1$ and $a \leq 1$.

\subsection{Achievable Secrecy Rate}

First, we give an achievable secrecy rate assuming that the transmitter and interferer use powers $P_1 \leq \bar{P}_1$ and
$P_2 \leq \bar{P}_2$, respectively.

\begin{theorem}\label{AchS}
When fixing the transmit power $(P_1,P_2)$ for the Gaussian WT-HI model given by (\ref{signalYZ}), the following secrecy rate is achievable:
\begin{eqnarray}\label{rate}
R_s(P_1, P_2) = \max\left\{R_s^{\mathrm{I}}(P_1,P_2),R_s^{\mathrm{II}}(P_1)\right\},
\end{eqnarray}
where $R_s^{\mathrm{I}}(P_1, P_2)$ is given by
\begin{align}
 R_s^{\mathrm{I}}(P_1, P_2) =
  \left\{ \begin{array}{ll}
  \gf\left(P_1\right)- \gf\left(\frac{aP_1}{1+P_2}\right)  &\mbox{if $b \geq 1+P_1$,}\\
  \gf\left(P_1+bP_2\right)- \gf\left(aP_1+P_2\right) &\mbox{if $1 \le  b < 1+P_1$,}\\
  \gf\left(\frac{P_1}{1+bP_2}\right)- \gf\left(\frac{aP_1}{1+P_2}\right)  &\mbox{if $b<1$,}
  \end{array} \nonumber \right.
\end{align}
and $R_s^{\mathrm{II}}(P_1)$ is given by
\begin{align}
 R_s^{\mathrm{II}}(P_1) = \left[\gf(P_1)-\gf(aP_1)\right]^{+},
\end{align}
with $\gf(x)\triangleq (1/2)\log(1+x)$.
\end{theorem}

\begin{proof}
This rate is achieved by using the coding scheme introduced in Section~\ref{sec:result}. The input distributions $\pi$ are chosen to be Gaussian $\mathcal{N}(0, P_1)$ and $\mathcal{N}(0, P_2)$ for $\Cc_1$ and $\Cc_2$, respectively. A sketch of a proof is provided in Appendix \ref{App-Gaussian-ach}.
\end{proof}


\subsubsection{Power Policy}\label{sec:power}

For the Gaussian WT-HI, power control plays an important role. Roughly speaking, the interferer may need to control its power so that it does not introduce too much interference to the primary transmission, while the transmitter may want to select its power so that the intended receiver is able to decode and cancel now helpful interference either fully or partially before decoding the primary transmission.

In the following, we give a power control strategy. We consider the cases when $a \geq 1$ and $a < 1$, separately.

When $a \geq 1$, we use the following transmit power:
\begin{eqnarray}\label{powerc1}
(P_1, P_2)=\left\{ \begin{array}{ll}
  (\min\{\bar{P}_1,P_1^{\ast}\}, \bar{P}_2)  &\mbox{if $b > 1, \bar{P}_2 > a-1 $,}\\
  (\bar{P}_1, \min\{\bar{P}_2, P_2^{\ast}\}) &\mbox{if $b < \frac{1}{a}, \bar{P}_2 > \frac{a-1}{1-ab}$, }\nonumber\\
  (0,0)  &\mbox{otherwise,}
  \end{array} \right.
\end{eqnarray}
and when $a < 1$, we use use the following transmit power:
\begin{eqnarray}\label{powerc2}
(P_1, P_2)=  &\left\{ \begin{array}{ll}
  (P_1^{\ast}, \bar{P}_2) &\mbox{if $b \geq \frac{1}{a}, \bar{P}_1 \geq b-1, \bar{P}_2 \geq \frac{1-a}{ab-1}$, }\\
  (\bar{P}_1,\min\{\bar{P}_2, P_2^{\ast}\}) &\mbox{if $b < 1, \bar{P}_1 \geq \frac{b-a}{a(1-b)}$, }\\
  (\bar{P}_1, 0)  &\mbox{if $1 \leq b \leq a^{-1}, \bar{P}_1 > \frac{b-1}{1-ab}$} \\
   ~& \mbox{or $a<b<1, \bar{P}_1 < \frac{b-a}{a(1-b)}$},\\
  (\bar{P}_1,\bar{P}_2)  &\mbox{otherwise,}
  \end{array} \right. \nonumber
\end{eqnarray}
where $P_1^{\ast}$ and $P_2^{\ast}$ are given by
\begin{align}
    P_1^{\ast}&=b-1,\label{past1}\\
    P_2^{\ast}&=\frac{(a-1)+\sqrt{(a-1)^2+(1-ab)\Delta}}{1-ab},\label{past2}\\
\mbox{and} \qquad   \Delta&=\frac{a}{b}(1+\bar{P}_1)-(1+a)\bar{P}_1.\label{pastdelta}
\end{align}

When $a>1$, a positive secrecy rate can be achieved when $b>1$ or $b \leq a^{-1}$ if the interferer's power $\bar{P}_2$ is large enough. When $b>1$, the interferer uses its full power $\bar{P}_2$ and the transmitter selects its power to guarantee that the intended receiver can first decode the interference (and cancel it). When $b<a^{-1}$, the intended receiver treats the interference as noise. In this case, the transmitter can use its full power $\bar{P}_1$ and the interferer controls its power (below $P_2^{\ast}$) to avoid excessive interference.

When $a<1$ and  $1 \leq b < a^{-1}$, the transmitter needs to restrict its power if it wants to let the receiver decode some interference. However, if the transmitter has a large power $\left(\bar{P}_1>\frac{b-1}{1-ab}\right)$, it is better to use all its power and to request that the interferer be silent. In the case when $a< b < 1$, the receiver treats the interference as noise. If the transmitter does not have enough power $\left(\bar{P}_1 <\frac{b-a}{a(1-b)}\right)$, the interference will hurt the intended receiver more than the eavesdropper.

\begin{lemma}\label{powerlemma}
The power policy maximizes the secrecy rate given in Theorem \ref{AchS}.
\end{lemma}

\begin{proof}
A proof is provided in Appendix \ref{powerlemmaproof}.
\end{proof}

\begin{remark}
The explicit form of the power policy gives some interesting insights into the achievable secrecy rate. For example, it is clear that an interference power $\bar{P}_2$ can benefit secrecy. In particular, when $\bar{P}_2$ is sufficiently large, a positive secrecy rate can be achieved when
\begin{equation}\label{psr_conds}
    \left( a < 1 \right) \mbox{~or~} \left(b >1\right) \mbox{~or~} \left( a>1 \mbox{~and~} b<\frac{1}{a}\right).
\end{equation}

In comparison, we recall that the secrecy capacity of the Gaussian wire-tap channel (when there is no interferer in the Gaussian WT-HI model) is
\begin{equation}\label{gwire-tap}
     R_s^{\mathrm{WT}}=\left[\gf(P_1)-\gf(aP_1)\right]^{+}
\end{equation}
and a positive secrecy rate can be achieved only when $a<1$.
\end{remark}

\subsubsection{Power-unconstrained Secrecy Rate}\label{powerun}

The secrecy rate achievable when the transmitter has unconstrained power depends only on the channel condition, and therefore is an important parameter of wire-tap-channel-based secrecy systems. Here, we refer to it as power-unconstrained secrecy rate. Note that the power-unconstrained secrecy capacity of the Gaussian wire-tap channel (assuming $a \neq 0$) is
\begin{equation}\label{limit1}
    \lim_{\bar{P}_1 \rightarrow \infty}\left[\gf(\bar{P}_1)-\gf(a\bar{P}_1)\right]^{+}=\frac{1}{2}\left[\log_{2}\frac{1}{a}\right]^{+}.
\end{equation}

The explicit form of the power policy facilitates
a limiting analysis, based on which we obtain the following result (assuming $ab\neq0$) for the power-unconstrained secrecy rate of the WT-HI model.
\begin{lemma}\label{poweruncon}
An achievable power unconstrained secrecy rate for the Gaussian WT-HI is
\begin{eqnarray}\label{limit2}
\lim_{\bar{P}_1,\bar{P}_2 \rightarrow \infty}R_s = \left\{
\begin{array}{ll}
  \frac{1}{2}\log_{2}b  &\mbox{if $b > \max(1,\frac{1}{a})$,}\\
  \frac{1}{2}\log_{2}\frac{1}{ab} &\mbox{if $b < \min(1,\frac{1}{a}$),}\\
  \frac{1}{2}\left[\log_{2}\frac{1}{a}\right]^{+}  &\mbox{otherwise.}
  \end{array} \right.
\end{eqnarray}
\end{lemma}

\begin{proof}
The proof can be found in Appendix \ref{powerunconproof}.
\end{proof}

\begin{remark}
Compared with the power-unconstrained rate without the help of interference, a gain of $(1/2)\log_{2}b$ can be observed for the WT-HI model when the interferer-receiver channel is good $\left(b > \max(1,\frac{1}{a})\right)$. Note that $(1/2)\log_{2}b$ is the power-unconstrained secrecy rate if the confidential message is sent from the interferer to the intended receiver in the presence of the eavesdropper. Therefore, this seems as if, `virtually', the message is given to the interferer secretly and that the interferer sends the message (a `cognitive' transmitter). This is particularly interesting because we do not assume that there is a secret transmitter-interferer channel (which would enable the interferer to relay the transmission). 

\end{remark}

\subsection{Upper Bounds}

Again, a simple upper bound on the secrecy capacity of the Gaussian WT-HI is the main channel capacity without a secrecy constraint. That is,
$$R_s \leq \gamma(\bar{P}_1).$$ In the following, we describe two additional upper bounds.

\subsubsection{Sato-type upper bound}

The first upper bound is based on the specialization of the Sato-type upper bound given by (\ref{sato}) to the Gaussian WT-HI model.

\begin{lemma}\label{lemma_gsb}
The secrecy capacity of the Gaussian WT-HI model given by (\ref{signalYZ}) is upper bounded as
\begin{equation}\label{satogb}
R_s \leq f(\bar{P}_1,\bar{P}_2,\rho^*(\bar{P}_1,\bar{P}_2)),
\end{equation}
where the function $f(P_1,P_2,\rho)$ is defined as
\begin{align} \label{GSatofunc}
f(P_1,P_2,\rho)=\frac{1}{2}\log{\frac{(1+P_1+bP_2)(1+aP_1+P_2)-(\rho+\sqrt{a}P_1+\sqrt{b}P_2)^2}{(1-\rho^2)(1+aP_1+P_2)}},
\end{align}
and $\rho^*(P_1,P_2)$ is given by
\begin{equation}\label{rhostar}
    \rho^{\star}(P_1,P_2)=\frac{(1+a)P_1+(1+b)P_2+(\sqrt{ab}-1)^2P_1P_2-\sqrt{\Theta}}{2(\sqrt{a}P_1+\sqrt{b}P_2)}
\end{equation}
with
\begin{align*}
    \Theta&=[(\sqrt{a}-1)^2P_1+(\sqrt{b}-1)^2P_2+(\sqrt{ab}-1)^2P_1P_2]\nonumber\\
          &\times[(\sqrt{a}+1)^2P_1+(\sqrt{b}+1)^2P_2+(\sqrt{ab}-1)^2P_1P_2].
\end{align*}
\end{lemma}

\begin{proof}
The proof can be found in Appendix \ref{gsatoproof}.
\end{proof}

\subsubsection{A Z-channel upper bound}

The second outer bound for the secrecy capacity of the Gaussian WT-HI model is motivated by \cite{Li:ISIT:08}.
To derive this bound, we assume that there is a genie to provide the interference codeword to the intended receiver.
In this case, the intended receiver can cancel interference without any cost and becomes interference-free ($b=0$).
The channel model becomes a one-sided interference channel (or Z-channel).

\begin{lemma}\label{lemma_zb}
The secrecy capacity of the Gaussian WT-HI model given by (\ref{signalYZ}) is upper bounded by
\begin{align}\label{zb_func}
  R_s &\leq  \frac{1}{2}\left[\log(1+\bar{P}_1)-\log(1+a\bar{P}_1)\right]^{+} + \frac{1}{2}\log\left[\frac{2(1+a\bar{P}_1)(1+\bar{P}_2)}{2+a\bar{P}_1+\bar{P}_2}\right].
\end{align}
\end{lemma}

\begin{proof}
The proof can be found in Appendix \ref{app_gzb}.
\end{proof}

\section{Numerical Examples}\label{sec:numerical}
\begin{figure}[ht]
  \centering
  \includegraphics[width=0.85\linewidth]{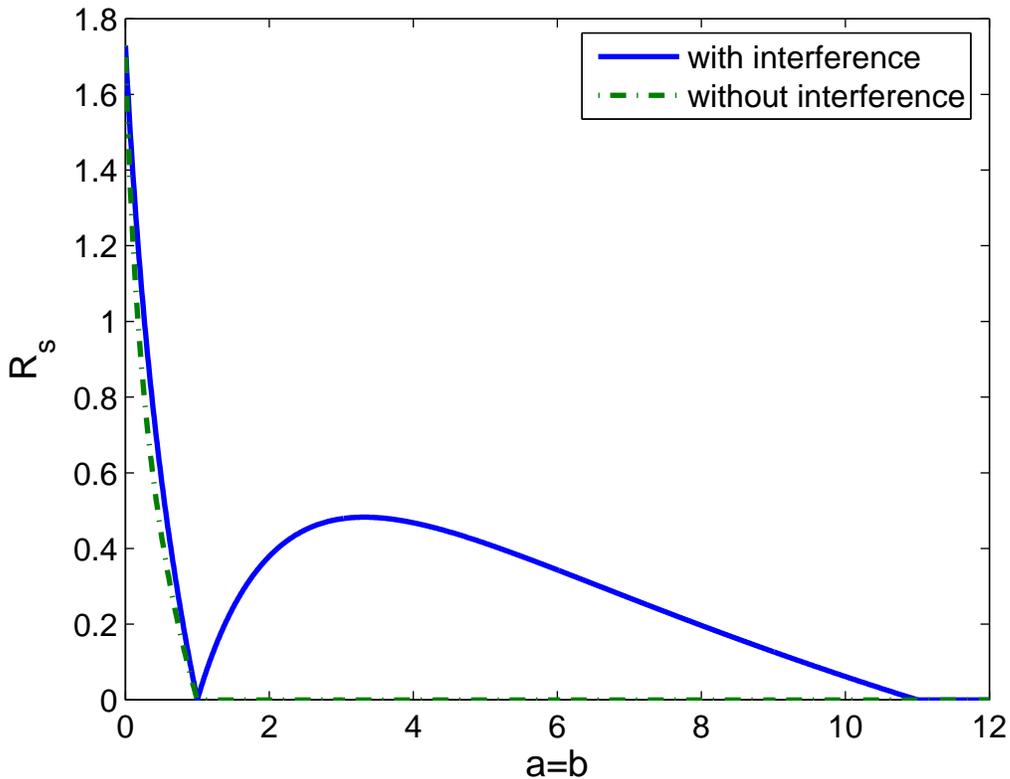}
  \caption{Achievable secrecy rate with or without the help of interference for a symmetric channel ($a=b$), where $\bar{P}_1=\bar{P}_2=10$.}\label{symRs}
  \vspace{-0.1in}
\end{figure}

In Fig. \ref{symRs}, the achievable secrecy rates with helping interference and without interference are shown for the symmetric Gaussian WT-HI model ($a=b$). In this example, the power constraints are $\bar{P}_1=\bar{P}_2=10$. The secrecy rate achieved with interference (here denoted by $R_s$) first decreases with $a$ when $a<1$; when $1<a \leq 3.26$, $R_s$ increases with $a$ because the intended receiver now can decode and cancel the interference, while the eavesdropper can only treat the interference as noise; when $a>3.26$, $R_s$ decreases again with $a$ because the interference does not affect the eavesdropper much when $a$ is large. It can also be found that nonzero secrecy rate can be achieved only when $a<1$ when there is no help of interference. However, nonzero secrecy rate can be achieved when $a<1+\bar{P}_2$ with the help of interference. It is clear that a larger value of $\bar{P}_2$ can improve the secrecy rate more. Hence, this result shows the value of exploiting interference to assist secrecy.

\begin{figure}[ht]
  \centering
  \includegraphics[width=0.85\linewidth]{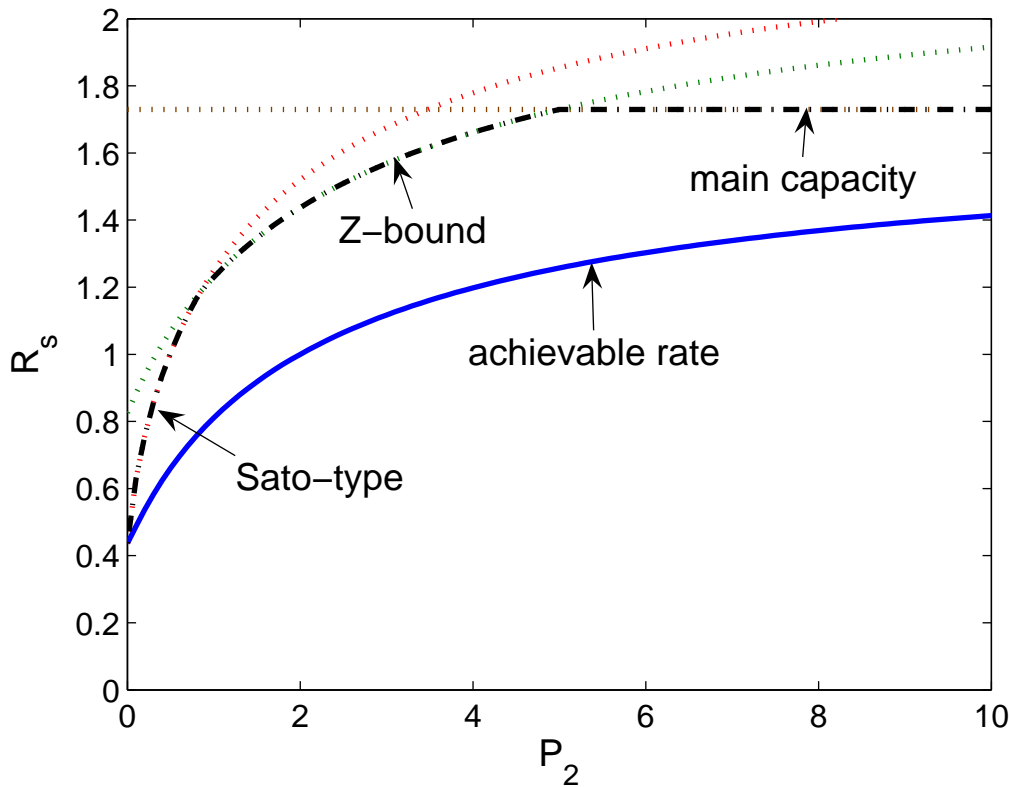}
  \caption{Achievable secrecy rate and upper bound versus $\bar{P}_2$, where $a=0.5$, $b=10$, and $\bar{P}_1=10$.}\label{RsP2_1}
\end{figure}
\begin{figure}[ht]
  \centering
  \includegraphics[width=0.85\linewidth]{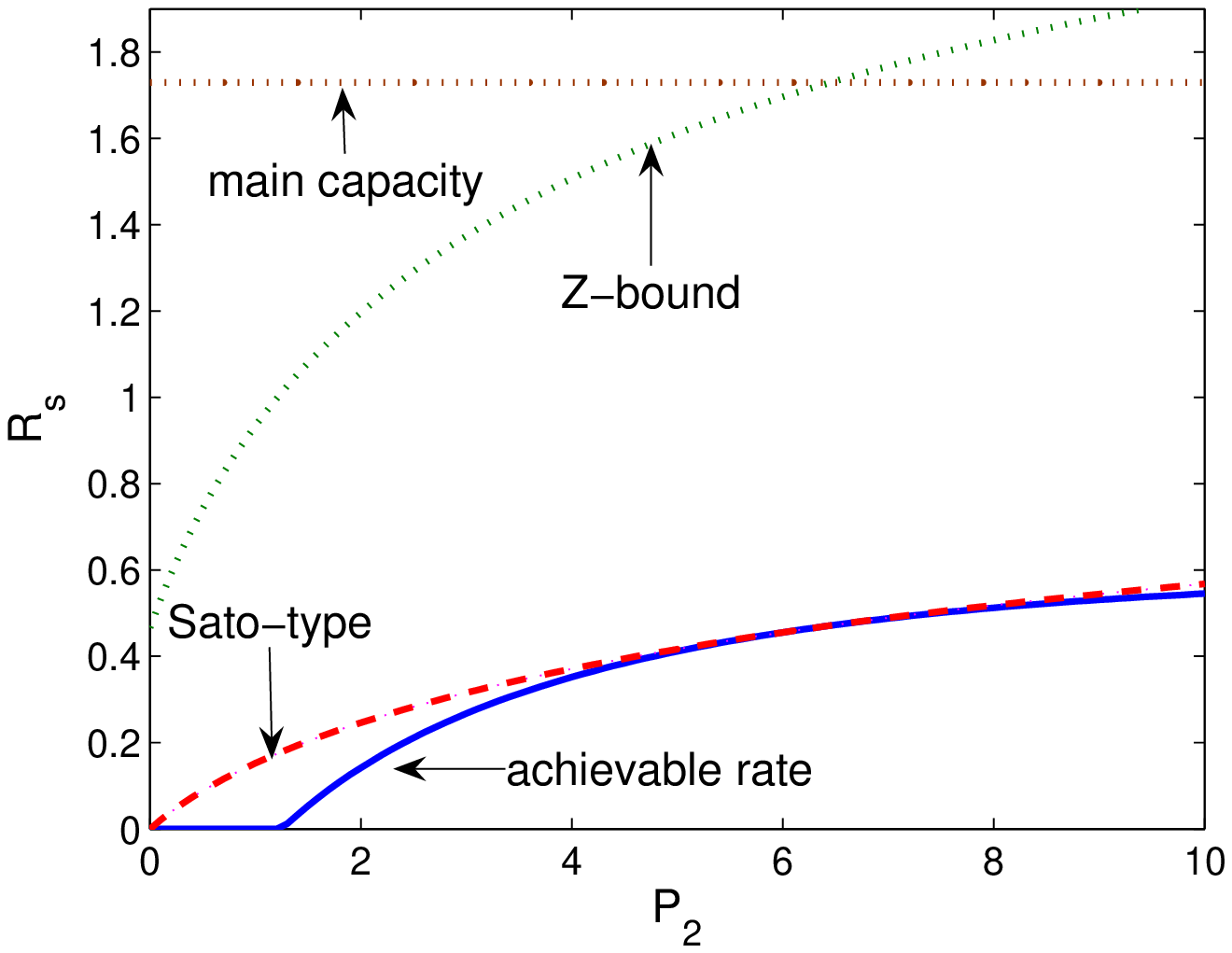}
  \caption{Achievable secrecy rate and upper bound versus $\bar{P}_2$, where $a=2$, $b=0.1$, and $\bar{P}_1=10$.}\label{RsP2_2}
\end{figure}
In Fig. \ref{RsP2_1} and Fig. \ref{RsP2_2}, we present numerical results to show the achievable rate and upper bounds versus $\bar{P}_2$ for some non-symmetric parameter settings of $a$ and $b$, where we again assume that $\bar{P}_1=10$. In Fig. \ref{RsP2_1}, $a$ and $b$ are fixed to be $0.5$ and $10$, respectively. Each of the three upper bounds is better than the others within some certain ranges of $P_2$. In particular, the Sato-type upper bound is the best when $\bar{P}_2$ is small, and the Z-channel bound becomes the best when $\bar{P}_2$ is relatively larger. It is also conceivable that when $\bar{P}_2$ is large (compared with a fixed $ \bar{P}_1$), the main channel capacity, though simple, is a good upper bound. In this case, the secret signals are hidden in very large interference at the eavesdropper, and secrecy can be achieved without sacrificing rate.

Also note that the Z-channel bound could be quite loose for some parameter settings of $a$ and $b$ (especially when $a>1$). As shown in Fig. \ref{RsP2_2}, where $a$ and $b$ are fixed to be $2$ and $0.1$, respectively, the Sato-type upper bound is uniformly better than the other two bounds during the shown range of $\bar{P}_2$. Our numerical results show that the Sato-type upper bound is relatively close to the achievable secrecy rate when $ab$ is close to 1. Note that $ab = 1$ corresponds to the degraded case, for which the Sato-type upper bound is always tight.


%

\section{Conclusions}\label{sec:conclusions}

In this paper, we have considered the use of the superposition property of the wireless medium to alleviate the eavesdropping issues caused by the broadcast nature of the medium. We have studied a wire-tap channel with a helping interferer (WT-HI), in which the interferer assists the secret communication by injecting independent interference. Our results show that interference, which seldom offers any advantage for problems not involving secrecy, can benefit secret wireless communication.

For general discrete memoryless WT-HI models, we have proposed an achievable secrecy rate which depends on the coding rate of the interference codebook. We have considered all possible interference coding rates. For a Gaussian WT-HI, we have given both the achievable secrecy rate and a power policy to optimize the secrecy rate. Our results show that the interferer can increase the secrecy level, and that a positive secrecy rate can be achieved even when the source-destination channel is worse than the source-eavesdropper channel. An important example of the Gaussian case is that in which the interferer has a better channel to the intended receiver than to the eavesdropper. Here, the interferer can send a (random) codeword at a rate that ensures that it can be decoded and subtracted from the received signal by the intended receiver, but cannot be decoded by the eavesdropper. Hence, only the eavesdropper is interfered with and the secrecy level of the confidential message can be increased. In addition, we have provided several computable upper bounds on the secrecy capacity of the Gaussian WT-HI. Each of the bounds can be a tighter upper bound under certain channel and power conditions. For some special cases, the upper bound is close to the achievable secrecy rate.

Future work of interest is to study the secrecy capacity of Gaussian interference channels with multiple confidential messages. The WT-HI model studied in this paper is a special case of two-user interference channel in which only one user has a confidential message to send. Therefore, we essentially have provided some results for studying the corner point of the secrecy capacity region of two-user interference channels, although a combination of the proposed achievable scheme and a time sharing strategy can provide an achievable secrecy rate region for general interference channels. We believe that the achievable scheme and upper bounds proposed in this paper can give in-depth insight and facilitate further study of general interference channel with multiple confidential messages.


\appendices

\section{Proof of Theorem~1}\label{achievability}

\begin{proof}
\subsubsection{Random Code Construction}

For a given distribution $p(x_1,x_2)=p(x_1)p(x_2)$, $\Cc_1$ and $\Cc_2$ are generated at random. More specifically, for codebook $\Cc_t$ ($t=1,2$), we generate $2^{nR_t}$ i.i.d. sequences each of length $n$ at random according to $p(\xv_t)=\prod_{i=1}^{n}p(x_{t,i})$.

A further step of codebook construction is the indexing of codewords for each codebook. Our proof here is based on an ``implicit" double binning technique.  That is, the $2^{nR_1}$ codewords in codebook $\Cc_1$ are randomly grouped into $2^{nR_{1,s}}$ bins each with $2^{nR_{1,d}}$ codewords, where $R_{1,d}=R_1-R_{1,s}$. Furthermore, in each bin, the $2^{nR_{1,d}}$ codewords are randomly grouped into $2^{nR'_{1,d}}$ sub-bins each with $2^{nR''_{1,d}}$ codewords (and thus $R_{1,d}=R'_{1,d}+R''_{1,d}$). Therefore, any codeword in $\Cc_1$ is indexed as $\xv_1(w_1,w'_1,w''_1)$ for $w_1 \in W_1 = \{1, \dots, 2^{nR_{1,s}}\}$, $w'_1 \in W'_1 = \{1, \dots, 2^{nR'_{1,d}}\}$ and $w''_1 \in W''_1 = \{1, \dots, 2^{nR''_{1,d}}\}$. The codewords in $\Cc_2$ are indexed as $\xv_2(w_2)$ for $w_2 \in W_2=\{1, \dots, 2^{nR_{2}}\}$.

\subsubsection{Encoding and Decoding}

In encoding, to send message $w_1 \in W_1$, the encoder at the transmitter selects $w'_1 \in W'_1$ and $w''_1 \in W''_1$ independently at random, and sends the codeword $\xv_1(w_1, w'_1, w''_1)$, while the encoder at the interferer selects $w_2 \in W_2$ at random and sends the codeword $\xv_2(w_2)$ to the channel.

In the decoding, after receiving $\yv_1$, the intended receiver declares that $\hat{w_1} \in W_1$ is received if
\begin{itemize}
  \item[(i)] (separate decoding): $\xv_1(\hat{w}_1, \hat{w}'_1, \hat{w}''_1)$ is the only codeword such that $\left<\xv_1(\hat{w}_1, \hat{w}'_1, \hat{w}''_1), \yv_1\right>$ is jointly typical; or
  \item [(ii)] (joint decoding): $\xv_1(\hat{w}_1, \hat{w}'_1, \hat{w}''_1)$ and $\xv_2(\hat{w}_2)$ are the only codeword pair such that $\left<\xv_1(\hat{w}_1, \hat{w}'_1, \hat{w}''_1), \xv_2(\hat{w}_2), \yv_1\right>$ is jointly typical.
\end{itemize}
The intended receiver makes an error if neither (i) nor (ii) occurs, or if $\hat{w}_1 \neq w_1$.

For any $(R_1, R_2) \in \left\{\Rc_1^{[\rm MAC]} \cup \Rc_1^{[\rm S]}\right\}$, the intended receiver can always decode the message $W_1$ reliably with an arbitrarily small error probability when $n$ is sufficiently large. Therefore, in the following, we only need to analyze the equivocation rate at the eavesdropper to account for the secrecy constraint.

The parameters $R'_{1,d}$ and $R''_{1,d}$ are different for each of two cases depending on the code rate $R_2$ of the (interference) codebook $\Cc_2$. Now we discuss those two cases separately. From the perspective of the eavesdropper, in case I, we consider $R_2 < I(X_2;Y_2|X_1)$, which corresponds to the rate region $\Rc_2^{[\rm MAC]}$; in case II, we consider $R_2 \geq I(X_2;Y_2|X_1)$, which corresponds to the rate region $\Rc_2^{[\rm S]}$.

\subsection{Case I: $R_2 < I(X_2;Y_2|X_1)$}

\subsubsection{Codebook Parameters}

We choose the following rate parameter for $R''_{1,d}$:
\begin{equation}\label{r1d_c1}
    R''_{1,d}=\min\left[ I(X_1,Y_2|X_2), I(X_1, X_2; Y_2) - R_2 \right] - \epsilon_1.
\end{equation}

\subsubsection{Equivocation Computation}

The equivocation at the eavesdropper is bounded as follows:
\begin{eqnarray}\label{eqv_comp_c1}
  H(W_1|\Yv_2)&\geq& H(W_1|\Yv_2, W'_1) \nonumber\\
  &=& H(W_1,\Yv_2| W'_1)-H(\Yv_2 | W'_1) \nonumber\\
  &=& H(W_1, \Yv_2, \Xv_1, \Xv_2|W'_1) - H(\Xv_1, \Xv_2| W_1, W'_1, \Yv_2) - H(\Yv_2| W'_1)\nonumber\\
  &=& H(\Xv_1, \Xv_2 |W'_1) + H(W_1, \Yv_2 | W'_1, \Xv_1, \Xv_2) - H(\Xv_1, \Xv_2 | W_1, W'_1, \Yv_2) - H(\Yv_2|W'_1)\nonumber\\
  &\geq& H(\Xv_1, \Xv_2 |W'_1) + H(\Yv_2 | \Xv_1, \Xv_2) - H(\Yv_2) - H(\Xv_1, \Xv_2|W_1,W'_1,\Yv_2)\nonumber\\
  &\geq& H(\Xv_1, \Xv_2 |W'_1) - I(\Xv_1, \Xv_2; \Yv_2) - H(\Xv_1, \Xv_2|W_1,W'_1, \Yv_2).
\end{eqnarray}

For the first term, we notice that
\begin{equation}\label{equv_T1_c1}
H(\Xv_1, \Xv_2|W'_1) = H(\Xv_1|W'_1) +H(\Xv_2) = n(R_{1,s}+R''_{1,d}+R_2).
\end{equation}

For the second term, we first have
\begin{equation*}\label{equv_T2-1_c1}
I(\Xv_1, \Xv_2; \Yv_2) \leq n\left[I(X_1, X_2 ;Y_2)-\delta_1\right],
\end{equation*}
where $\delta_1 \rightarrow 0$ as $n \rightarrow \infty$. We also have
\begin{eqnarray*}\label{equv_T2-2_c1}
I(\Xv_1, \Xv_2; \Yv_2) &=& I(\Xv_2; \Yv_2) + I(\Xv_1; \Yv_2 |\Xv_2) \nonumber \\
&\leq& H(\Xv_2) + I(\Xv_1; \Yv_2 |\Xv_2) \nonumber\\
&\leq& nR_2 + n\left[I(X_1, Y_2 |X_2)-\delta_1\right].
\end{eqnarray*}
Therefore, we have
\begin{equation}\label{equv_T2_c1}
I(\Xv_1, \Xv_2; \Yv_2) \leq \min\left[ I(X_1,X_2; Y_2), I(X_1; Y_2 |X_2) + R_2 \right] - \delta_1 = n\left(R''_{1,d} + R_2\right)-\delta_1,
\end{equation}
where the last equality is based on (\ref{r1d_c1}).

To bound the third term, we consider the (joint) decoding of $W''_1$ and $W_2$ at the eavesdropper assuming that $W_1$ and $W'_1$ are given to the eavesdropper as side information. Given that $W_1=w_1$ and $W'_1=w'_1$, we assume that $w''_1$ and $w_2$ are sent. The eavesdropper declares that $\xv_1(w_1, w'_1, \hat{w}''_1 )$ and $\xv_2(w_2)$ are sent if $\xv_1(w_1, w'_1, \hat{w}''_1 )$ is the only codeword in the sub-bin $\mathcal{B}(w_1, w'_1)$ and $\xv_2(\hat{w}_2)$ is the only codeword in $\Cc_2$, such that $\left\langle \xv_1(w_1, w'_1, \hat{w}''_1 ), \xv_2(\hat{w}_2), \yv_2 \right\rangle $ is jointly typical. The eavesdropper makes an error if $(\hat{w}''_1, \hat{w}_2) \neq (w''_1, w_2)$ or if there is no such a codeword pair jointly typical with $\yv_2$. According to (\ref{r1d_c1}), the rate pair $(R''_{1,d}, R_2)$ satisfies the following condition:
\begin{eqnarray*}\label{R2mac}
\left\{
\begin{array}{ll}
  R''_{1,d} \leq I(X_1; Y_2 | X_2),\\
  R_2 \leq I(X_2, Y_2 |X_1),\\
  R''_{1,d} + R_2 \leq I(X_1,X_2;Y_2).
  \end{array} \right.
\end{eqnarray*}
Therefore, the probability of error is arbitrarily small when $n$ is large. Based on Fano's inequality, we have
\begin{equation*}
H(\Xv_1, \Xv_2|W_1=w_1,W'_1=w'_1, \Yv_2)\leq n\delta_2.
\end{equation*}
Hence, we have
\begin{equation}\label{equv_T3_c1}
H(\Xv_1, \Xv_2 | W_1,W'_1,\Yv_2) = \sum_{w_1, w'_1}p(w_1,w'_1)H(\Xv_1, \Xv_2|W_1=w_1,W'_1=w'_1,\Yv_2) \leq n\delta_2.
\end{equation}
By combining (\ref{eqv_comp_c1}) with (\ref{equv_T1_c1}), (\ref{equv_T2_c1}) and (\ref{equv_T3_c1}), we have
\begin{equation*}
H(W_1|\Yv_2) \geq n\left[R_{1,s}+(\delta_1-\epsilon_1-\delta_2)\right]=n\left(R_{1,s}-\epsilon\right),
\end{equation*}
where $\epsilon \rightarrow 0$ as $n \rightarrow \infty$. Therefore, the secrecy constraint is satified.

\subsection{Case II: $R_2 > I(X_2;Y_2|X_1)$}

\subsubsection{Codebook Parameters}

We choose the following rate parameter for $R''_{1,d}$:
\begin{equation}\label{r1d_c2}
    R''_{1,d}=I(X_1; Y_2) - \epsilon_1.
\end{equation}

Note that in the encoding, the interfering encoder selects $w_2 \in W_2$ at random and sends the codeword $\xv_2(w_2)$ to the channel. In order to prove for Case II, we assume that this is done as in the following procedure. We suppose that the $2^{nR_2}$ codewords in codebook $\Cc_2$ are randomly grouped into $2^{nR'_{2}}$ bins each with $2^{nR''_{2}}$ codewords, where $R''_{2}=R_2-R'_{2}$. Therefore, any codeword in $\Cc_2$ is indexed as $\xv_2(w'_2,w''_2)$ for $w'_2 \in W'_2 = \{1, \dots, 2^{nR'_{2}}\}$ and $w''_2 \in W''_2 = \{1, \dots, 2^{nR''_{2}}\}$. During encoding, the encoder at the helper selects $w'_2 \in W'_2$ and $w''_2 \in W''_2$ independently at random, and sends the codeword $\xv_2(w'_2,w''_2)$. This is equivalent to that a random codeword $\xv_2(w_2)$ ($w_2 = w'_2 \times w''_2$) is sent. To facilitate the proof, we let
\begin{equation}\label{r2d_c2}
   R''_2=I(X_2;Y_2|X_1)-\epsilon_2.
\end{equation}

\subsubsection{Equivocation Computation}

Following steps similar to those as given by (\ref{eqv_comp_c1}), the equivocation at the eavesdropper is bounded by:
\begin{eqnarray}\label{eqv_comp_c2}
  H(W_1|\Yv_2)&\geq& H(W_1|\Yv_2, W'_1, W'_2) \nonumber\\
  &\geq& H(\Xv_1, \Xv_2 |W'_1, W'_2) - I(\Xv_1, \Xv_2; \Yv_2) - H(\Xv_1, \Xv_2|W_1,W'_1,W'_2,\Yv_2).
\end{eqnarray}
For the first term, we have that
\begin{align}\label{equv_T1_c2}
H(\Xv_1, \Xv_2|W'_1, W'_2) &= H(\Xv_1|W'_1) +H(\Xv_2|W'_2) \nonumber\\
&= n(R_{1,s}+R''_{1,d})+nR''_2 = n\left[R_{1,s}+I(X_1,X_2;Y_2)-\epsilon_3\right],
\end{align}
where $\epsilon_3=\epsilon_1+\epsilon_2 \rightarrow 0$ as $n \rightarrow \infty$.
For the second term, we have that
\begin{equation}\label{equv_T2_c2}
I(\Xv_1, \Xv_2; \Yv_2) \leq n\left[I(X_1, X_2 ;Y_2)-\delta_1\right].
\end{equation}
To bound the third term, we consider the (joint) decoding of $W''_1$ and $W''_2$ at the eavesdropper assuming that $W_1$, $W'_1$ and $W'_2$ are given to the eavesdropper as side information. For the rate pair $(R''_{1,d}, R''_2)=(I(X_1; Y_2) - \epsilon_1, I(X_2; Y_2 | X_2) - \epsilon_2)$, we can show that the probability of error is arbitrarily small when $n$ is large. Hence, we also have
\begin{equation}\label{equv_T3_c2}
H(\Xv_1, \Xv_2 | W_1,W'_1,W'_2, \Yv_2) \leq n\delta_2.
\end{equation}
By combining (\ref{eqv_comp_c2}) with (\ref{equv_T1_c2}), (\ref{equv_T2_c2}) and (\ref{equv_T3_c2}), we have
\begin{equation*}
H(W_1|\Yv_2) \geq n\left[R_{1,s}+(\delta_1-\epsilon_3-\delta_2)\right]=n\left(R_{1,s}-\epsilon\right),
\end{equation*}
where $\epsilon = \epsilon_3 + \delta_2 - \delta_1 \rightarrow 0$ as $n \rightarrow \infty$. Therefore, the secrecy constraint is also satified for case II.
\end{proof}

\section{Proof of Theorem \ref{Souterbound}}\label{App-Souter}

\begin{proof}
The secrecy requirement implies that
\begin{equation}\label{Asecrecy}
    nR_s=H(W_1) \leq H(W_1|Y_2^n)+n\epsilon,
\end{equation}
and Fano's inequality implies that
\begin{equation}\label{AFano}
    H(W_1|Y_1^n) \leq n\epsilon R_1+ h(\epsilon) \triangleq n\delta.
\end{equation}
Based on (\ref{Asecrecy}) and (\ref{AFano}), we have
\begin{align}
    nR_s &\leq H(W_1|Y_2^n)+n\epsilon \nonumber\\
         &\leq H(W_1|Y_2^n) - H(W_1|Y_1^n) + n(\epsilon + \delta) \nonumber \\
         &\leq H(W_1|Y_2^n) - H(W_1|Y_1^n, Y_2^n) + n(\epsilon + \delta)\label{A1-1}\\
         &= I(W_1;Y_1^n|Y_2^n)+ n(\epsilon + \delta)\nonumber\\
         &\leq I(X_1^n, X_2^n;Y_1^n|Y_2^n)+n(\epsilon + \delta) \label{A1-2}\\
         &\leq \sum_{i=1}^{n}I(X_{1,i}, X_{2,i};Y_{1,i}|Y_{2,i})+n(\epsilon + \delta), \label{A1-3}
\end{align}
where (\ref{A1-1}) is due to the fact that conditioning reduces entropy, and
(\ref{A1-2}) follows since $W_1 \rightarrow (X_1^n, X_2^n) \rightarrow (Y_1^n,
Y_2^n)$ forms a Markov chain.

Now, it is observed that the secrecy capacity of the WT-HI depends only on the marginal distributions $P_{Y_1|X_1,X_2}$ and $P_{Y_2|X_1,X_2}$, and not on any further structure of the joint distribution $P_{Y_1,Y_2|X_1,X_2}$. This can be easily proved because the average error probability $P_e$ defined by (\ref{pe}) depends on the marginal distribution $P_{Y_1|X_1,X_2}$ only, and the equivocation rate $H(W_1|Y_2^n)/n$ depends on the marginal distribution $P_{Y_2|X_1,X_2}$ only. Hence, the secrecy capacity is the same for any channel described by (\ref{eqalmar}) whose marginal distributions are the same. We can replace $(Y_1,Y_2)$ with $(\tilde{Y}_1,\tilde{Y}_2)$ defined by (\ref{eqalmar}) and obtain (\ref{sato}).
\end{proof}

\section{Proof of Theorem~\ref{AchS}}
\label{App-Gaussian-ach}

\begin{proof}
The achievability is based on the coding scheme introduced in Section~\ref{sec:result}, with
the input distributions $\pi$ chosen to be Gaussian $\mathcal{N}(0, P_1)$ and $\mathcal{N}(0, P_2)$ for $\Cc_1$ and $\Cc_2$, respectively.
Here, we discuss only the coding parameters $R_1$, $R_2$ and $R_{1,d}$ in Theorem~\ref{thm:WT-HI}.

When $a \geq 1+P_2$, which is the very strong eavesdropping case as discussed in Section~\ref{sec:result}, we have $R_s=0$.

Next, we discuss $R_s^{\mathrm{I}}(P_1,P_2)$ for the case when $a \leq 1+P_2$. Here, we choose $R_2=\gf(P_2)$ and $R_{1,d}=\gf\left(\frac{aP_1}{1+P_2}\right)$.
\begin{enumerate}
  \item When $b \geq 1+P_1$, we have $R_2 \leq \gf\left(\frac{bP_2}{1+P_1}\right)$. In this case, we choose $R_1=\gf(P_1)$. The intended receiver can first perform a separate decoding using $\Cc_2$ and cancel interference, and then
can decode at the rate of $R_1=\gf(P_1)$ using $\Cc_1$ (virtually a clean channel). The secrecy rate is $R_s=R_1-R_{1,d}=\gf(P_1)-\gf\left(\frac{aP_1}{1+P_2}\right)$.

  \item When $1 \leq b \leq 1+P_1$, we have $\gf\left(\frac{bP_2}{1+P_1}\right) \leq R_2 \leq \gf(bP_2)$. In this case, we choose $R_1=\gf(P_1+bP_2)-\gf(P_2)$.
The intended receiver performs joint decoding using $\Cc_1$ and $\Cc_2$. The secrecy rate is $R_s=R_1-R_{1,d}=\left[\gf(P_1+bP_2)-\gf(P_2)\right]-\gf\left(\frac{aP_1}{1+P_2}\right)=\gf(P_1+bP_2)-\gf(aP_1+P_2)$.

  \item When $b \leq 1$, we have $\gf(P_2) \geq \gf(bP_2)$. In this case, we choose $R_1=\gf\left(\frac{P_1}{1+bP_2}\right)$. The intended receiver performs a separate decoding using $\Cc_1$.
The secrecy rate is $R_s=R_1-R_{1,d}=\gf\left(\frac{P_1}{1+bP_2}\right)-\gf\left(\frac{aP_1}{1+P_2}\right)$.
\end{enumerate}

Under certain conditions, to choose $R_2=0$ and $R_{1,d}=\gf(aP_1)$ can yield a higher secrecy rate. In this case, the secrecy rate is $R_s^{\mathrm{II}}(P_1)=\left[\gf(P_1)-\gf(aP_1)\right]^{+}$. Therefore, $\max\left(R_s^{\mathrm{I}},R_s^{\mathrm{II}}\right)$ can be achieved.
\end{proof}

\section{Proof of Lemma ~\ref{powerlemma}}
\label{powerlemmaproof}

\begin{proof}
First, we notice that $R_s^{\mathrm{II}}$ can be viewed as a special result of $R_s^{\mathrm{I}}$ if power is optimized (by applying power $(P_1,P_2)=(\bar{P}_1,0)$).
Hence, to optimize $R_s$ given by (\ref{rate}), we can ignore $R_s^{\mathrm{II}}$ and consider only the optimization of $R_s^{\mathrm{I}}$ with respect to $P_1$ and $P_2$.

For convenience, we denote
\begin{align*}
    R_{s1}&=\gf(P_1)-\gf\left(\frac{aP_1}{1+P_2}\right),\\
    R_{s2}&=\gf(P_1+bP_2)-\gf(aP_1+P_2),\\
\mbox{and}~\qquad  R_{s3}&=\gf\left(\frac{P_1}{1+bP_2}\right)-\gf\left(\frac{aP_1}{1+P_2}\right),
\end{align*}
which are all functions of $P_1$ and $P_2$. Depending on the channel parameters $(a,b)$ and power $(P_1, P_2)$, only one of the three functions is active.
When $0 \leq P_1 \leq \bar{P}_1$ and $0 \leq P_2 \leq \bar{P}_2$, all three functions are bounded. Therefore, there always exists a global maximum, which might be a maximum of one certain function or at a cross point of two functions. The subsections below is to search for the maximum point by checking the gradients and comparing with the boundary points.

Since we care about the sign of the (partial) derivatives of these functions,
for convenience, we say that two real numbers $x$ and $y$ satisfy $x \backsim y$ if they have the same sign. It can be shown that we have
\begin{align*}
    &\quad \frac{\partial R_{s1}}{\partial P_1} \backsim 1-a+P_2, \quad
    \frac{\partial R_{s2}}{\partial P_1} \backsim \frac{\partial R_{s3}}{\partial P_1} \backsim 1-a + (1-ab)P_2,\\
    &\quad \frac{\partial R_{s1}}{\partial P_2} \geq 0, \quad \frac{\partial R_{s2}}{\partial P_2} \backsim b-1+(ab-1)P_1,\\
    &\mbox{and~~} \frac{\partial R_{s3}}{\partial P_2} \backsim \left[b(ab-1)P_2^2+2b(a-1)P_2+a-b+a(1-b)P_1\right]P_1,
\end{align*}
and all cases are discussed in the following subsections.

\subsection{$a > 1$}

For the case when $a > 1$, we consider $b > 1$, $a^{-1} < b \leq 1$, and $b \leq a^{-1}$, respectively.

\begin{enumerate}
\item[1)] $b > 1$:

\begin{enumerate}
  \item [i)] $\bar{P}_{2} \leq a-1$: we cannot obtain a positive secrecy rate, and therefore, $(P_1,P_2)=(0,0)$.

  \item [ii)] $\bar{P}_{2} > a-1$: we choose $(P_1,P_2)=(\min(\bar{P}_1,P_1^{*}),\bar{P}_2)$ because of the following:

  If $\bar{P}_1 \leq b-1$, $R_{s1}$ is active,  $(\frac{\partial R_{s1}}{\partial P_1} \geq 0, \frac{\partial R_{s1}}{\partial P_2} \geq 0)$ and therefore $(P_1,P_2)=(\bar{P}_1,\bar{P}_2)$. Now if $\bar{P}_1 > b-1$, once we choose $P_1 > P_1^{*}$, $R_{s2}$ is active and $(\frac{\partial R_{s2}}{\partial P_1} \leq 0, \frac{\partial R_{s1}}{\partial P_2} \geq 0)$. This forces us to set $(P_1,P_2)=(P_1^{*},\bar{P}_2)$.
\end{enumerate}

\item[2)] $a^{-1} < b \leq 1$: $R_{s3}$ is active, and we have $(\frac{\partial R_{s3}}{\partial P_1} \leq 0, \frac{\partial R_{s3}}{\partial P_2} \leq 0)$. Therefore, $(P_1,P_2)=(0,0)$.

\item[3)] $b \leq a^{-1}$:  $R_{s3}$ is active.

\begin{enumerate}
\item[i)] $\bar{P}_{2} \leq \frac{a-1}{1-ab}$: $(\frac{\partial R_{s3}}{\partial P_1} \leq 0, \frac{\partial R_{s3}}{\partial P_2} \leq 0)$ and therefore, $(P_1,P_2)=(0,0)$.

  \item[ii)] $\bar{P}_{2} > \frac{a-1}{1-ab}$: we choose $(P_1,P_2)=(\bar{P}_1,\min(\bar{P}_2,P_2^{*}))$ because of the following

  If $\frac{a-1}{1-ab} < \bar{P}_2 \leq P_2^{*}$, $(\frac{\partial R_{s3}}{\partial P_1} \geq 0, \frac{\partial R_{s3}}{\partial P_2} \geq 0)$ and therefore $(P_1,P_2)=(\bar{P}_1,\bar{P}_2)$. Now if $\bar{P}_2 > P_2^{*}$, once we choose $P_2 > P_2^{*}$, $(\frac{\partial R_{s3}}{\partial P_1} \geq 0, \frac{\partial R_{s3}}{\partial P_2} \leq 0)$. This forces us to set $(P_1,P_2)=(\bar{P}_1,P_2^{*})$.
\end{enumerate}
\end{enumerate}

\subsection{$a \leq 1$}
For the case when $a \leq 1$, we consider $b > a^{-1}$, $1 < b \leq a^{-1}$, and $b \leq 1$, respectively.

\begin{enumerate}
\item[1)]$b > a^{-1}$:
\begin{enumerate}
\item[i)] $\bar{P}_{1} \leq b-1$: $R_{s1}$ is active and we have $(\frac{\partial R_{s1}}{\partial P_1} \geq 0, \frac{\partial R_{s1}}{\partial P_2} \geq 0)$. Therefore, we have $(P_1,P_2)=(\bar{P}_2,\bar{P}_2)$.

  \item[ii)] $\bar{P}_{1} > b-1$: If $\bar{P}_2 \leq \frac{1-a}{ab-1}$, $R_{s2}$ is active, $(\frac{\partial R_{s2}}{\partial P_1} \geq 0, \frac{\partial R_{s2}}{\partial P_2} \geq 0)$ and therefore we choose $(P_1,P_2)=(\bar{P}_1,\bar{P}_2)$. Now if $\bar{P}_2 > \frac{1-a}{ab-1}$, once we choose $P_2 > \frac{1-a}{ab-1}$, $(\frac{\partial R_{s2}}{\partial P_1} \leq 0, \frac{\partial R_{s1}}{\partial P_2} \geq 0)$. This forces us to choose $(P_1,P_2)=(P_1^{*},\bar{P}_2)$. After comparing with the rate achieved by using $(P_1,P_2)=(\bar{P}_1,\frac{1-a}{ab-1})$, we find that $(P_1,P_2)=(P_1^{*},\bar{P}_2)$ is better.

\end{enumerate}

\item[2)] $1 < b \leq a^{-1}$:

\begin{enumerate}
\item[i)] $\bar{P}_{1} \leq \frac{b-1}{1-ab}$:  when $\bar{P}_{1} \leq b-1$, $R_{s1}$ is active and we have $(\frac{\partial R_{s1}}{\partial P_1} \geq 0, \frac{\partial R_{s1}}{\partial P_2} \geq 0)$; when $b-1 < \bar{P}_{1} \leq \frac{b-1}{1-ab}$, $R_{s2}$ is active and we have $(\frac{\partial R_{s2}}{\partial P_1} \geq 0, \frac{\partial R_{s2}}{\partial P_2} \geq 0)$. Therefore, we choose $(P_1,P_2)=(\bar{P}_1,\bar{P}_2)$.

  \item[ii)] $\bar{P}_{1} > \frac{b-1}{1-ab}$: we choose $(P_1,P_2)=(\bar{P}_1,0)$ because of the following:

 If one chooses $P_1 > \frac{b-1}{1-ab}$, $R_{s2}$ is active and $(\frac{\partial R_{s2}}{\partial P_1} \geq 0, \frac{\partial R_{s2}}{\partial P_2} \leq 0)$, therefore we need $(P_1,P_2)=(\bar{P}_1,0)$. After comparing with $(P_1,P_2)=(\frac{b-1}{1-ab},\bar{P}_2)$, we find that $(P_1,P_2)=(\bar{P}_1,0)$ is better.
\end{enumerate}

\item[3)] $b \leq 1$: $R_{s3}$ is active.

\begin{enumerate}
  \item[i)] $\bar{P}_{1} \leq \frac{b-a}{a(1-b)}$: since $(\frac{\partial R_{s3}}{\partial P_1} \geq 0, \frac{\partial R_{s3}}{\partial P_2} \leq 0)$, we choose $(P_1,P_2)=(\bar{P}_1,0)$.

  \item[ii)] $\bar{P}_{1} > \frac{b-a}{a(1-b)}$: we choose $(P_1,P_2)=(\bar{P}_1,\min(\bar{P}_2,P_2^{*}))$ because of the following:

  If $\bar{P}_2 \leq P_2^*$, $(\frac{\partial R_{s3}}{\partial P_1} \geq 0, \frac{\partial R_{s3}}{\partial P_2} \geq 0)$, and we choose $(P_1,P_2)=(\bar{P}_1,\bar{P}_2)$.
  If $\bar{P}_2 > P_2^*$ and once choosing $P_2 > P_2^*$, $(\frac{\partial R_{s3}}{\partial P_1} \geq 0, \frac{\partial R_{s3}}{\partial P_2} \leq 0)$. This forces us to choose $(P_1,P_2)=(P_1,P_2^*)$.
\end{enumerate}
\end{enumerate}

\end{proof}

\section{Proof of Lemma \ref{poweruncon}}
\label{powerunconproof}

\begin{proof}
When $b > \max(1,a^{-1})$, the power policy uses $(P_1,P_2)=(P_1^{*},\bar{P}_2)$, where $P_1^{*}=b-1$.
Therefore, the achievable secrecy rate is
\begin{align*}
    R_s&=\gf\left(P_1^{*}\right)-\gf\left(\frac{aP_1^*}{1+\bar{P}_2}\right)\\
    &=\frac{1}{2}\log b - \frac{1}{2} \log\left(1+\frac{a(b-1)}{1+\bar{P}_2}\right).
\end{align*}
After taking the limit with respect of $\bar{P}_2$, we have
$R_s  =\frac{1}{2}\log b$.

When $b < \min(1,a^{-1})$, the power policy uses $(P_1,P_2)=(\bar{P}_1,P_2^*)$, where $P_2^*$ is given by (\ref{past2}) and can be shown to be
\begin{equation*}
    P_2^*=\sqrt{\frac{(a-b-ab)\bar{P}_1}{b(1-ab)}},
\end{equation*}
when $\bar{P}_1$ is large. The achievable secrecy rate is
\begin{align*}
    R_s&=\gf\left(\frac{\bar{P}_1}{1+bP_2^*}\right)-\gf\left(\frac{a\bar{P}_1}{1+P_2^{*}}\right).
\end{align*}
After taking the limit with respect of $\bar{P}_1$, we have $R_s =\frac{1}{2}\log \frac{1}{ab}$.

For other cases, the power policy uses $P_2=0$ and the achievable secrecy rate remains at $R_s=\frac{1}{2}\left[\log_{2}\frac{1}{a}\right]^{+}.$
\end{proof}

\section{Proof of Lemma (\ref{lemma_gsb})}
\label{gsatoproof}

\begin{proof}
To use the result given by (\ref{sato}), we let
\begin{align}\label{VVMD}
    &\tilde{Y}_1 = X_1 + \sqrt{b} X_2 + \tilde{Z}_1\\
 \mbox{~and~} \qquad &\tilde{Y}_2 = \sqrt{a} X_1 + X_2 +
\tilde{Z}_2,
\end{align}
where $\tilde{Z}_1$ and $\tilde{Z}_2$ are arbitrarily correlated Gaussian random variables with zero-means and unit variances. We also let
$\rho$ denote the covariance between $\tilde{Z}_1$ and $\tilde{Z}_2$, i.e.,
\begin{equation*}
    \mathrm{Cov}(\tilde{Z}_1,\tilde{Z}_2) = \rho.
\end{equation*}
It can be observed that $P_{Y_1,Y_2|X_1,X_2}$ and $P_{\tilde{Y}_1,\tilde{Y}_2|X_1,X_2}$ have the same marginal distribution and satisfy the condition given by (\ref{eqalmar}).

Note that $I(X_1,X_2;\tilde{Y}_1|\tilde{Y}_2)$ is a function of the transmit powers $P_1$ and $P_2$, and the noise covariance $\rho$. Hence, we denote it by
\begin{align} \label{ffunc}
I(X_1,X_2;\tilde{Y}_1|\tilde{Y}_2)=f(P_1,P_2,\rho).
\end{align}

Now we show that $f(P_1,P_2,\rho)$ can be evaluated by (\ref{GSatofunc}). To show this, $I(X_1,X_2;\tilde{Y}_1|\tilde{Y}_2)$ is evaluated as
\begin{align}
    &I(X_1,X_2;\tilde{Y}_1|\tilde{Y}_2) \nonumber\\
         &= I(X_1,X_2;\tilde{Y}_1,\tilde{Y}_2)-I(X_1,X_2;\tilde{Y}_2) \nonumber \\
         &= [H(\tilde{Y}_1,\tilde{Y}_2)-H(\tilde{Y}_1,\tilde{Y}_2|X_1,X_2)]-[h(\tilde{Y}_2)-h(\tilde{Y}_2|X_1,X_2)]\nonumber\\
         &= h(\tilde{Y}_1|\tilde{Y}_2)-h(\tilde{Z}_1|\tilde{Z}_2)\nonumber\\
         &= h(\tilde{Y}_1|\tilde{Y}_2)- \frac{1}{2}\log[2\pi e(1-\rho^2)].
\end{align}
For convenience, we let
\begin{equation}\label{tc}
    t=\frac{E[\tilde{Y}_1\tilde{Y}_2]}{E[\tilde{Y}_2^2]}.
\end{equation}
We have
\begin{align}
    h(\tilde{Y}_1|\tilde{Y}_2) &= h(\tilde{Y}_1- t\tilde{Y}_2|\tilde{Y}_2) \nonumber \\
         &\leq h(\tilde{Y}_1- t\tilde{Y}_2) \label{A2-1}\\
         &\leq \frac{1}{2}\log[2\pi e \mathrm{Var}(\tilde{Y}_1-
         t\tilde{Y}_2)]\label{A2-2},
\end{align}
where (\ref{A2-2}) follows from the maximum-entropy theorem and both
equalities in (\ref{A2-1}) and (\ref{A2-2}) hold true when
$(X_1,X_2)$ are Gaussian.

Furthermore, we have
\begin{align*}
\mathrm{Var}(\tilde{Y}_1-t\tilde{Y}_2)=
1+P_1+bP_2-\frac{(\rho+\sqrt{a}P_1+\sqrt{b}P_2)^2}{1+aP_1+P_2}.
\end{align*}
Hence, $I(X_1,X_2;\tilde{Y}_1|\tilde{Y}_2)$ can be evaluated by
(\ref{GSatofunc}).

It is easy to verify that $f(P_1,P_2,\rho)$ is an increasing function of both $P_1$ and $P_2$ for any given $\rho$, and $f(P_1,P_2,\rho)$ is a convex function of $\rho$ for any given $P_1$ and $P_2$. It can be shown that when $P_1$ and $P_2$ are given, the minimum of $f(P_1,P_2,\rho)$ occurs when $\rho$ is chosen to be $\rho^{\star}$ given by (\ref{rhostar}).

Therefore, the Sato-type upper bound can be calculated as
\begin{equation*}
    \min_{\rho}\max_{(P_1,P_2)}f(P_1,P_2,\rho)=f(\bar{P}_1,\bar{P}_2,\rho^*(\bar{P}_1,\bar{P}_2)).
\end{equation*}
\end{proof}

\section{Proof of Lemma \ref{lemma_zb} }
\label{app_gzb}

\begin{proof}
Based on the secrecy requirement given by (\ref{Asecrecy}) and Fano's inequality given by (\ref{AFano}), we have
\begin{align}
    nR_s &\leq H(W_1|Y_2^n) - H(W_1|Y_1^n) + n(\epsilon + \delta) \nonumber \\
         &\leq I(W_1;Y_1^n) - I(W_1;Y_2^n) + n(\epsilon + \delta). \nonumber
\end{align}
For simplicity, we omit the term $n(\epsilon + \delta)$ since it does not change the result. Now we let
\begin{align}\label{V1V2}
    &V_1^n=X_1^n+Z_1^n\\
 \mbox{~and~} \qquad &V_2^n=\sqrt{a}X_1^n+Z_2^n,
\end{align}
and proceed with the following steps:
\begin{align}
    nR_s &\leq I(W_1;Y_1^n, V_1^n) - I(W_1;Y_2^n,V_2^n) + I(W_1; V_2^n|Y_2^n) \nonumber \\
         &=I(W_1;V_1^n) + I(W_1;Y_1^n | V_1^n) - I(W_1; V_2^n) - I(W_1;Y_2^n|V_2^n) + I(W_1; V_2^n|Y_2^n) \nonumber \\
         &=I(W_1;V_1^n) - I(W_1; V_2^n) + I(W_1; V_2^n|Y_2^n),\label{zb-t1}
\end{align}
where we use the fact that $I(W_1;Y_1^n|V_1^n)=0$ and $I(W_1;Y_2^n|V_2^n)=0$ since each of $W_1 \leftrightarrow V_1^n \leftrightarrow Y_1^n$ and $W_1 \leftrightarrow V_2^n \leftrightarrow Y_2^n$ forms a Markov chain. We therefore have
\begin{equation}\label{zb-t2}
    nR_s \leq \left[I(W_1;V_1^n) - I(W_1; V_2^n)\right]^{+} + I(W_1; V_2^n|Y_2^n).
\end{equation}

Based on the result for the Gaussian wire-tap channel \cite{Leung-Yan-Cheong:IT:78}, we have
\begin{equation}\label{zb-t3}
   \left[I(W_1;V_1^n) - I(W_1; V_2^n)\right]^{+}  \leq \frac{n}{2}\left[\log(1+\bar{P}_1)-\log(1+a\bar{P}_1)\right]^{+}.
\end{equation}

Now, we bound $I(W_1; V_2^n|Y_2^n)$ via the following steps:
\begin{align}
    I(W_1;V_2^n|Y_2^n) &\leq I(W_1,X_1^n;V_2^n| Y_2^n) \nonumber \\
         &=I(X_1^n;V_2^n| Y_2^n) \nonumber \\
         &=h(V_2^n|Y_2^n)-h(V_2^n|X_1^n,Y_2^n) \nonumber \\
         &=\left[h(Y_2^n,V_2^n)-h(Y_2^n,V_2^n|X_1^n)\right]-h(Y_2^n)+h(Y_2^n|X_1^n) \notag\\
         &= I(X_1^n; Y_2^n, V_2^n)-h(Y_2^n)+h(Y_2^n|X_1^n). \notag
\end{align}
Since
\begin{equation*}
I(X_1^n; Y_2^n, V_2^n)=I(X_1^n;V_2^n)=h(V_2^n)-h(Z_2^n),
\end{equation*}
we have
\begin{align}\label{zb-t4}
  I(W;V_2^n|Y_2^n) &\leq h(V_2^n)+h(Y_2^n|X_1^n)-h(Y_2^n)-h(Z_2^n) \notag\\
  &=h(\sqrt{a}X_1^n+Z_2^n) + h(X_2^n+Z_2^n) - h(\sqrt{a}X_1^n+X_2^n+Z_2^n)-h(Z_2^n).
\end{align}

Since we assume that $X_1^n$ and $X_2^n$ are independent (and both are also independent of $Z_2^n$), based on the subset sum entropy power inequality (EPI) \cite{Madiman:IT:07}, we have
\begin{align}
  \exp\left(\frac{2}{n}h(\sqrt{a}X_1^n+X_2^n+Z_2^n)\right) \geq \frac{1}{2}\left[\exp\left(\frac{2}{n}h(\sqrt{a}X_1^n+Z_2^n)\right)+\exp\left(\frac{2}{n}h(X_2^n+Z_2^n)\right) \right].
\end{align}
By letting $t_1=h(\sqrt{a}X_1^n+Z_2^n)$ and $t_2=h(X_2^n+Z_2^n)$, we have
\begin{align}\label{zb-t5}
  h(\sqrt{a}X_1^n+X_2^n+Z_2^n) \geq \frac{n}{2}\left\{\log\left[\exp\left(\frac{2t_1}{n}\right)+\exp\left(\frac{2t_2}{n}\right)\right]-\log{2}\right\}.
\end{align}
Using (\ref{zb-t5}) in (\ref{zb-t4}), we obtain
\begin{align}
  I(W;V_2^n|Y_2^n) &\leq t_1 + t_2 -  \frac{n}{2}\left\{\log\left[\exp\left(\frac{2t_1}{n}\right)+\exp\left(\frac{2t_2}{n}\right)\right]-\log{2}\right\} -  \frac{n}{2}\log (2 \pi e)\notag \\
  &=\frac{n}{2}\log\left[\frac{\exp\left(\frac{2(t_1+t_2)}{n}\right)}{\exp\left(\frac{2t_1}{n}\right)+\exp\left(\frac{2t_2}{n}\right)}\right]-\frac{n}{2} \log(\pi e).\label{zb-t6}
\end{align}
Note that the bound given by (\ref{zb-t6}) is an increasing function of both $t_1$ and $t_2$. From the maximum-entropy theorem, we have
\begin{align*}
  t_1 &\leq \frac{n}{2}\log\left(2\pi e (1+a\bar{P}_1)\right),\\
  t_2 &\leq \frac{n}{2}\log\left(2\pi e (1+\bar{P}_2)\right),
\end{align*}
where the equalities hold when both $X_1^n$ and $X_2^n$ are i.i.d. Gaussian.
Therefore,
\begin{equation}\label{zb-t7}
  I(W_1;V_2^n|Y_2^n) \leq \frac{n}{2}\log\left[\frac{2(1+a\bar{P}_1)(1+\bar{P}_2)}{2+a\bar{P}_1+\bar{P}_2}\right].
\end{equation}

Finally, by combining (\ref{zb-t2}), (\ref{zb-t3}), and (\ref{zb-t7}), we obtain the upper bound given by (\ref{zb_func}).
\end{proof}

\bibliographystyle{IEEEtran}
\bibliography{MacFC,IFC}

\begin{thebibliography}{10}
\providecommand{\url}[1]{#1}
\csname url@samestyle\endcsname
\providecommand{\newblock}{\relax}
\providecommand{\bibinfo}[2]{#2}
\providecommand{\BIBentrySTDinterwordspacing}{\spaceskip=0pt\relax}
\providecommand{\BIBentryALTinterwordstretchfactor}{4}
\providecommand{\BIBentryALTinterwordspacing}{\spaceskip=\fontdimen2\font plus
\BIBentryALTinterwordstretchfactor\fontdimen3\font minus
  \fontdimen4\font\relax}
\providecommand{\BIBforeignlanguage}[2]{{%
\expandafter\ifx\csname l@#1\endcsname\relax
\typeout{** WARNING: IEEEtran.bst: No hyphenation pattern has been}%
\typeout{** loaded for the language `#1'. Using the pattern for}%
\typeout{** the default language instead.}%
\else
\language=\csname l@#1\endcsname
\fi
#2}}
\providecommand{\BIBdecl}{\relax}
\BIBdecl

\bibitem{Wyner:BSTJ:75}
A.~D. Wyner, ``The wire-tap channel,'' \emph{Bell Syst. Tech. J.}, vol.~54,
  no.~8, pp. 1355--1387, Oct. 1975.

\bibitem{Csiszar:IT:78}
I.~Csisz{\'{a}}r and J.~K{\"{o}}rner, ``Broadcast channels with confidential
  messages,'' \emph{IEEE Trans. Inf. Theory}, vol.~24, no.~3, pp. 339--348, May
  1978.

\bibitem{Leung-Yan-Cheong:IT:78}
S.~K. Leung-Yan-Cheong and M.~Hellman, ``The {G}aussian wire-tap channel,''
  \emph{IEEE Trans. Inf. Theory}, vol.~24, no.~4, pp. 451--456, July 1978.

\bibitem{Carleial:IT:75}
A.~Carleial, ``A case where interference does not reduce capacity,'' \emph{IEEE
  Trans. Inf. Theory}, vol.~21, no.~5, pp. 569--570, Sep. 1975.

\bibitem{Sato:IT:81}
H.~Sato, ``The capacity of the {G}aussian interference channel under strong
  interference,'' \emph{IEEE Trans. Inf. Theory}, vol.~27, no.~6, pp. 786--788,
  Nov. 1981.

\bibitem{Han:IT:81}
T.~Han and K.~Kobayashi, ``A new achievable rate region for the interference
  channel,'' \emph{IEEE Trans. Inf. Theory}, vol.~27, no.~1, pp. 49--60, Jan.
  1981.

\bibitem{Sato:IT:78}
H.~Sato, ``On degraded {G}aussian two-user channels,'' \emph{IEEE Trans. Inf.
  Theory}, vol.~24, no.~5, p. 634–640, Sep. 1978.

\bibitem{Costa:IT:85}
M.~H.~M. Costa, ``On the {G}aussian interference channel,'' \emph{IEEE Trans.
  Inf. Theory}, vol.~31, no.~5, pp. 607--615, Sep. 1985.

\bibitem{Kramer:IT:04}
G.~Kramer, ``Outer bounds on the capacity of gaussian interference channels,''
  \emph{IEEE Trans. Inf. Theory}, vol.~50, no.~3, p. 581–586, Mar. 2004.

\bibitem{Etkin:IT:08}
R.~H. Etkin, D.~N.~C. Tse, and H.~Wang, ``Gaussian interference channel
  capacity to within one bit,'' \emph{IEEE Trans. Inf. Theory}, vol.~54,
  no.~12, p. 5534–5562, Dec. 2008.

\bibitem{Motahari:IT:09}
A.~S. Motahari and A.~K. Khandni, ``Capacity bounds for the {G}aussian
  interference channel,'' \emph{IEEE Trans. Inf. Theory}, vol.~55, no.~2, pp.
  620--643, Feb. 2009.

\bibitem{Annapureddy:ITA:08}
V.~Annapureddy and V.~Veeravalli, ``Sum capacity of the {G}aussian interference
  channel in the low interference regime,'' in \emph{Proc. ITA Workshop}, La
  Jolla, CA, Jan. 2008.

\bibitem{Shang:IT:09}
X.~Shang, G.~Kramer, and B.~Chen, ``A new outer bound and the
  noisy-interference sum-rate capacity for {G}aussian interference channels,''
  \emph{IEEE Trans. Inf. Theory}, vol.~55, no.~2, pp. 689--699, Feb. 2009.

\bibitem{Liu:IT:08}
R.~Liu, I.~Maric, P.~Spasojevi\'{c}, and R.~Yates, ``Discrete memoryless
  interference and broadcast channels with confidential messages: Secrecy
  capacity regions,'' \emph{IEEE Trans. Inf. Theory}, vol.~54, no.~6, pp.
  2493--2507, Jun. 2008.

\bibitem{Liang:IT:09}
Y.~Liang, A.~Somekh-Baruch, H.~V. Poor, S.~{Shamai (Shitz)}, and S.~Verd\'{u},
  ``Capacity of cognitive interference channels with and without secrecy,''
  \emph{IEEE Trans. Inf. Theory}, vol.~55, no.~2, pp. 604--619, Feb. 2009.

\bibitem{Li:ISIT:08}
Z.~Li, R.~D. Yates, and W.~Trappe, ``Secrecy capacity region of a class of
  one-sided interference channel,'' in \emph{Proc. IEEE Int. Symp. Inf.
  Theory}, Totronto, Canada, Jul. 2008, pp. 379--383.

\bibitem{Yates:ISIT:08}
R.~Yates, D.~Tse, and Z.~Li, ``Secret communication on interference channels,''
  in \emph{Proc. IEEE Int. Symp. Inf. Theory}, Totronto, Canada, Jul. 2008, pp.
  374--378.

\bibitem{Koyluoglu:IT:08}
O.~Koyluoglu, H.~{El Gamal}, L.~Lai, and H.~V. Poor, ``Interference alignment
  for secrecy,'' \emph{submitted to IEEE Trans. Inf. Theory}, 2008.

\bibitem{Liang:IT:06}
Y.~Liang and H.~V. Poor, ``Multiple access channels with confidential
  messages,'' \emph{IEEE Trans. Inf. Theory}, vol.~54, no.~3, pp. 976--1002,
  Mar. 2008.

\bibitem{Liu:ISIT:06}
R.~Liu, I.~Maric, R.~Yates, and P.~Spasojevi\'{c}, ``The discrete memoryless
  multiple access channel with confidential messages,'' in \emph{Proc.\ IEEE
  Int. Symp. Information Theory}, Seattle, WA, USA, July 2006.

\bibitem{Tekin:IT:07}
E.~Tekin and A.~Yener, ``The general {G}aussian multiple-access and two-way
  wire-tap channels: Achievable rates and cooperative jamming,'' \emph{IEEE
  Trans. Inf. Theory}, vol.~54, no.~6, pp. 2735--2751, Jun. 2008.

\bibitem{Lai:IT:06}
L.~Lai and H.~{El~Gamal}, ``The relay-eavesdropper channel: Cooperation for
  secrecy,'' \emph{IEEE Trans. Inf. Theory}, vol.~54, no.~9, pp. 4005--4019,
  Sep. 2008.

\bibitem{Yusel:CISS:07}
M.~Yuksel and E.~Erkip, ``The relay channel with a wire-tapper,'' in
  \emph{Proc.\ 41st Annual Conference on Information Sciences and Systems},
  Baltimore, MD, Mar. 2007.

\bibitem{Liang:ITS:08}
Y.~Liang, H.~V. Poor, and S.~{Shamai (Shitz)}, ``Information theoretic
  security,'' \emph{Foundations and Trends in Communications and Information
  Theory}, vol.~5, no. 4-5, pp. 355--580, 2008.

\bibitem{Madiman:IT:07}
M.~Madiman and A.~Barron, ``Generalized entropy power inequalities and
  monotonicity properties of information,'' \emph{IEEE Trans. Inf. Theory},
  vol.~53, no.~7, pp. 2317--2329, Jul. 2007.

\end{thebibliography}

\end{document}